\documentclass[preprint, notrackchanges]{aastex631} 
\usepackage{multirow, url, color}
\usepackage{bbding, amssymb} 
\usepackage{ulem} 
\usepackage{verbatim} 
\usepackage{enumerate}

\usepackage{amsmath}
\usepackage{booktabs} 
\usepackage{graphicx, subfigure}
\usepackage{appendix}

\def\gsim{\;\lower4pt\hbox{${\buildrel\displaystyle >\over\sim}$}\;}
\def\lsim{\;\lower4pt\hbox{${\buildrel\displaystyle <\over\sim}$}\;}
\def\grls{\;\lower4pt\hbox{${\buildrel\displaystyle >\over <}$}\;}

\newcommand{\dblind}[1]{{\color[rgb]{0,0,0} will be revealed after the double-blind review process.}}

\shorttitle{A statistical study on SEP spectra}
\shortauthors{Wang \& Guo}
\graphicspath{{./}{figures/}}

\begin{document}

\title{A statistical study on the peak and fluence spectra of Solar Energetic Particles observed over 4 solar cycles}

\author[0009-0007-7848-1501]{Yubao Wang}
\affiliation{Deep Space Exploration Laboratory/School of Earth and Space Sciences, University of Science and Technology of China, Hefei 230026, China}

\author[0000-0002-8707-076X]{Jingnan Guo}
\affiliation{Deep Space Exploration Laboratory/School of Earth and Space Sciences, University of Science and Technology of China, Hefei 230026, China}
\affiliation{CAS Center for Excellence in Comparative Planetology, University of Science and Technology of China, Hefei 230026, China}
\affiliation{Collaborative Innovation Center of Astronautical Science and Technology, Harbin 150001, China}

\correspondingauthor{Jingnan Guo}
\email{jnguo@ustc.edu.cn}

\begin{abstract}
{Solar energetic particles (SEPs) are an important space radiation source, especially for the space weather environment in the inner heliosphere. The energy spectrum of SEP events is crucial both for evaluating their radiation effects and for understanding their acceleration process at the source region and their propagation mechanism.
In this work, we investigate the properties of the SEP peak flux spectra and the fluence spectra and their potential formation mechanisms using statistical methods.
We aim to advance our understanding of both SEPs' acceleration and propagation mechanisms.
Employing the dataset of ESA's Solar Energetic Particle Environment Modelling (SEPEM) program, we have obtained and fitted the peak-flux and fluence proton spectra of more than a hundred SEP events from 1974 to 2018. We analyzed the relationship among the solar activity, X-ray peak intensity of solar flares and the SEP spectral parameters.
Based on the assumption that the initial spectrum of accelerated SEPs generally has a power-law distribution and also the diffusion coefficient has a power-law dependence on particle energy, we can assess both the source and propagation properties using the observed SEP event peak flux and fluence energy spectra.  We confirm that SEPs' spectral properties are influenced by the solar source and the interplanetary conditions and their transportation process can be influenced by different phases of solar cycle.
This study provides an observational perspective on the double power-law spectral characteristics of the SEP energy spectra, showing their correlation with the adiabatic cooling and diffusion processes during the particle propagation from the Sun to the observer. This contributes to a deeper understanding of the acceleration and propagation of SEP events, in particular the possible origins of the double-power law.}
\end{abstract}


\section{Introduction}\label{sec:intro}
In order to carry out space exploration activities, we need to have a clear understanding of the radiation environment in space. The energy spectrum of high-energy particles in space is a direct input for evaluating radiation dose. As the ultimate source, the Sun not only releases high-energy charged particles, but also significantly affects the propagation of particles. So the observed energy spectrum varies due to the different properties of particle acceleration and propagation processes. Various studies have attempted to use the observed data and existing knowledge to reconstruct these two aspects of information \citep[see, e.g.,][and references therein]{klein2017acceleration}.
\par Typically, there are two mechanisms that contribute to the acceleration of charged particles \citep{reames1999particle,Reames2013}: magnetic reconnection acceleration associated with flares and shock wave acceleration associated with Coronal Mass Ejections (CMEs). Ideally, we may be able to distinguish different acceleration processes by comparing the composition and state of particles in the chromosphere and corona, such as proton-to-electron ratios, $^3\mathrm{He}$ compositions, and ion charge states \citep{mason2007,klein2017acceleration}.  But due to the fact that a SEP event is often associated with the onsets of both a solar flare and a CME, it is difficult to distinguish the role of the two types of acceleration processes. Furthermore, tracking the source of SEP events becomes more challenging when considering the propagation process of particles \citep[see, e.g.,][and references therein]{guo2024particle}.
\par Once SEPs are released into interplanetary space, their propagation can be theoretically described by the Fokker-Planck equation below \citep{parker1965}:
\begin{align}
\frac{\partial U}{\partial t}&=-\frac{V_{sw}}{r^{2}} \frac{\partial}{\partial r}\left(r^{2} U\right)+\frac{2 V_{sw}}{3 r} \frac{\partial}{\partial E}[UEn(E)]+\frac{1}{r^{2}} \frac{\partial}{\partial r}\left(\kappa r^{2} \frac{\partial U}{\partial r}\right);\label{eqn:Fokker-Planck} \\
n(E)&=\frac{pdE}{Edp}=\frac{E+2m_0c^2}{E+m_0c^2}.
\end{align}
Here $U(r,E,t)$ represents the probability distribution over kinetic energy $E$ for location $r$ and time $t$; $V_{sw}$ is solar wind speed; $\kappa$ is diffusion coefficient; and $m_0$ is rest mass of the particle. $n(E)$ represents the normalized conversion coefficient between particle momentum $p$ and kinetic energy $E$. For relativistic particles, $n(E)\approx 1$, and for non-relativistic particles, $n(E)\approx 2$. The terms on the right-hand side of Eq. \ref{eqn:Fokker-Planck} originate from three particle transport mechanisms: convection, adiabatic cooling, and diffusion.  Each of these transport mechanisms may modify the original acceleration characteristics of energetic particles.
\par In this study, we focus on the properties of SEP proton energy spectra and statistically investigate the correlation between solar activity index and the properties of SEP events. We aim to extract information about the acceleration and propagation processes of SEPs combining analytical descriptions of particle propagation and parameters obtained from observations. Analytical details about SEP energy spectra will be introduced in Sect. \ref{sec:SEP energy spectra}. The dataset, SEP event list, and fitting method will be introduced in Sect. \ref{sec:SEPEM dataset and Methods}.  Subsequently, we will examine the distributions of SEP proton energy spectra parameters, the relationships between integral fluence spectra and peak flux spectra, and how they change with the solar cycle in Sect. \ref{sec:result}. The results and conclusions will be summarized in Sect. \ref{sec:summary}.

\section{Analytical descriptions of SEP energy spectra}\label{sec:SEP energy spectra}
The SEP energy spectrum is the flux distribution of SEPs versus the particle energy, which is determined by both the acceleration process and transport effects. The SEP spectrum can be directly measured by high-energy particle detectors in space, and some large SEP events can also be indirectly recorded by ground neutron detectors or muon detectors when the secondary particles have energy high enough to reach the surface of Earth, also known as Ground-Level enhancement (GLE) events \citep[see, e.g., Sect. 3 of][and references therein]{guo2024particle}. By adopting different integration time windows, we can study the event energy spectrum and its time evolution. The most-frequently referred one is the integrated fluence spectrum of the entire event. This spectrum accumulates the flux of particles over different energies from the beginning to the end of the event, and it is crucial for evaluating the total radiation effect of the event. Another important one is the time-of-maximum (TOM) energy spectrum or peak flux spectrum formed by the maximum flux at each energy bin. This method can minimize the impact of solar wind on energetic particle propagation processes, such as diffusion, convection, and velocity dispersion effects, allowing for a more reasonable estimation of the energy spectrum of the source region \citep{Hollebeke1975}.

\subsection{Integral fluence spectra}\label{sec:Integral fluence spectra}
Previous researchers have proposed multiple functions to describe the integral fluence spectra of SEP events. The most commonly used functions are the single power law with an exponential rollover function \citep[e.g.,][]{Tylka2001} and the double power law function \citep[e.g.,][]{mewaldt2012,Raukunen2018}. The latter is also called the Band function, which was originally proposed for fitting $\gamma$-ray burst spectra \citep{band1993batse}. It is confirmed that the Band formulation generally fits SEP integral fluence spectra better, especially for those large events  \cite [e.g.,][]{mewaldt2005}. Therefore we choose this function to fit SEP integral fluence spectra in this study. The Band function is given by:
\begin{equation}\label{eqn:Band_function}
 \frac{d J}{d E}=\left\{\begin{array}{ll}
&C E^{\gamma_{1}} \exp \left(-E / E_{0}\right) ,\\& \text { if } E <\left(\gamma_{1}-\gamma_{2}\right) E_{0}; \\
&C E^{\gamma_{2}}\left\{\left[\left(\gamma_{1}-\gamma_{2}\right) E_{0}\right]^{\left(\gamma_{1}-\gamma_{2}\right)} \exp \left(\gamma_{2}-\gamma_{1}\right)\right\}, \\& \text { if } E \geq\left(\gamma_{1}-\gamma_{2}\right) E_{0}.
\end{array}\right.
\end{equation}
Here C is an overall fluence normalization coefficient in the same units of $dJ/dE$; $\gamma_1$ is the power-law index at the low-energy range; $\gamma_2$ is the power-law index at the high-energy range; $E$ is particle's kinetic energy as defined in Eq. \ref{eqn:Fokker-Planck}; $E$ and $E_0$ are measured in energy/nucleon; $(\gamma_1-\gamma_2)E_0$ is break-point energy which separates the low-energy and high-energy ranges. Both the Band function and its first derivative are continuous at this break point. But we have to note that there is no clear physical meaning of the four free parameters of Band function, and this function is still a semi-phenomenological model. Due to the mixing of propagation effects such as diffusion and adiabatic cooling, it is difficult to directly obtain information on the source and propagation process of high-energy particles through the parameters defined by the integrated energy spectra. More discussions will be given in Sects. \ref{sec:result} and \ref{sec:summary}.

\subsection{Peak flux spectra}\label{sec:Peak flux spectra}
When neglecting convection and adiabatic deceleration terms, the transportation of energetic particles can be treated as a process of simple time-dependent spherical diffusion from a point source. Then the time profile of a well-connected SEP event flux can be described by a simple function \cite[e.g.,][]{parker1963interplanetary}:
\begin{align}
j(p, r, t)=\frac{N_{0}(p)}{4 \pi p^{2}} \frac{\exp \left(-r^{2} / 4 t \kappa(p)\right)}{2 \sqrt{\pi}(t \kappa(p))^{3 / 2}}, \label{eqn:flux_time_profile}
\end{align}
where $j(p, r, t)$ is particle's flux at location $r$, time $t$ with momentum $p$,  $\kappa (p)$ is energy-dependent diffusion coefficient, $N_{0}(p)$ is the number of particles per unit momentum released at the Sun. From Eq. \ref{eqn:flux_time_profile}, one can derive the peak flux spectrum for $t=t_{max}=r^2/(6\kappa(p))$, and one can see that peak flux spectrum has the same shape with the source spectrum \citep{forman1986acceleration}:
\begin{align}
j\left(p, r, t_{max}(r, p)\right)=\frac{(6 / e)^{3 / 2}}{2 \sqrt{\pi} r^{3}} \frac{N_{0}(p)}{4 \pi p^{2}}\propto \frac{N_{0}(p)}{4 \pi p^{2}}. \label{eqn:peak_flux_spectrum_analytical}
\end{align}
\par For the early stages of SEP events, the convective effect of particle propagation can be ignored compared to the diffusion effect \cite[e.g.,][]{mccracken1971decay}. When considering adiabatic deceleration  and assuming a simple power-law diffusion coefficient 
\begin{align}
\kappa=\kappa_0 E^{\alpha},\label{eqn:diffusion coefficient}
\end{align}
\cite{kurt1981} estimated the particle's adiabatic energy loss $\Delta E$ as:
\begin{align}
\Delta E=E_s-E=E_s-\left(E_s^{\alpha}-\frac{4}{9}\frac{ V_{sw}r}{\kappa_0}\right)^{\frac{1}{\alpha}},\label{eqn:adiabatic_energy_loss}
\end{align}
where $E_s$ is particle's kinetic energy directly being released at the Sun, $E$ is particle's kinetic energy at location $r$, $V_{sw}$ is solar wind speed, $\kappa_0$ is normalization coefficient in the same unit of diffusion coefficient $\kappa$, $\alpha$ is power-law index.
\par Previous studies have shown that for the two acceleration processes of SEPs, it is relatively reasonable to assume that the energy spectrum after acceleration follows a simple power-law \cite[e.g.,][]{jones1991,dierckxsens2015,fu2006}:
\begin{align}
j(E_s)=j_0E_s^{-\gamma}, \label{eqn:source_spectrum}
\end{align}
where $j_0$ is normalization coefficient in the same units of particle's flux $j$, $\gamma$ is the index of the power-law energy spectrum after acceleration. Based on the conservation of particle number and combined with Eq. \ref{eqn:adiabatic_energy_loss}, the particle energy spectrum at location $r$ can be obtained as:
\begin{align}
j(r,E)=j_0\left( E^{\alpha}+\frac{4}{9}\frac{V_{sw}}{\kappa_0}r\right)^{-\frac{\gamma}{\alpha}}. \label{eqn:peak_spectrum}
\end{align}
\par
\begin{figure*}
\begin{minipage}[t]{1.0\linewidth}
\centering
\includegraphics[width=0.95\hsize]{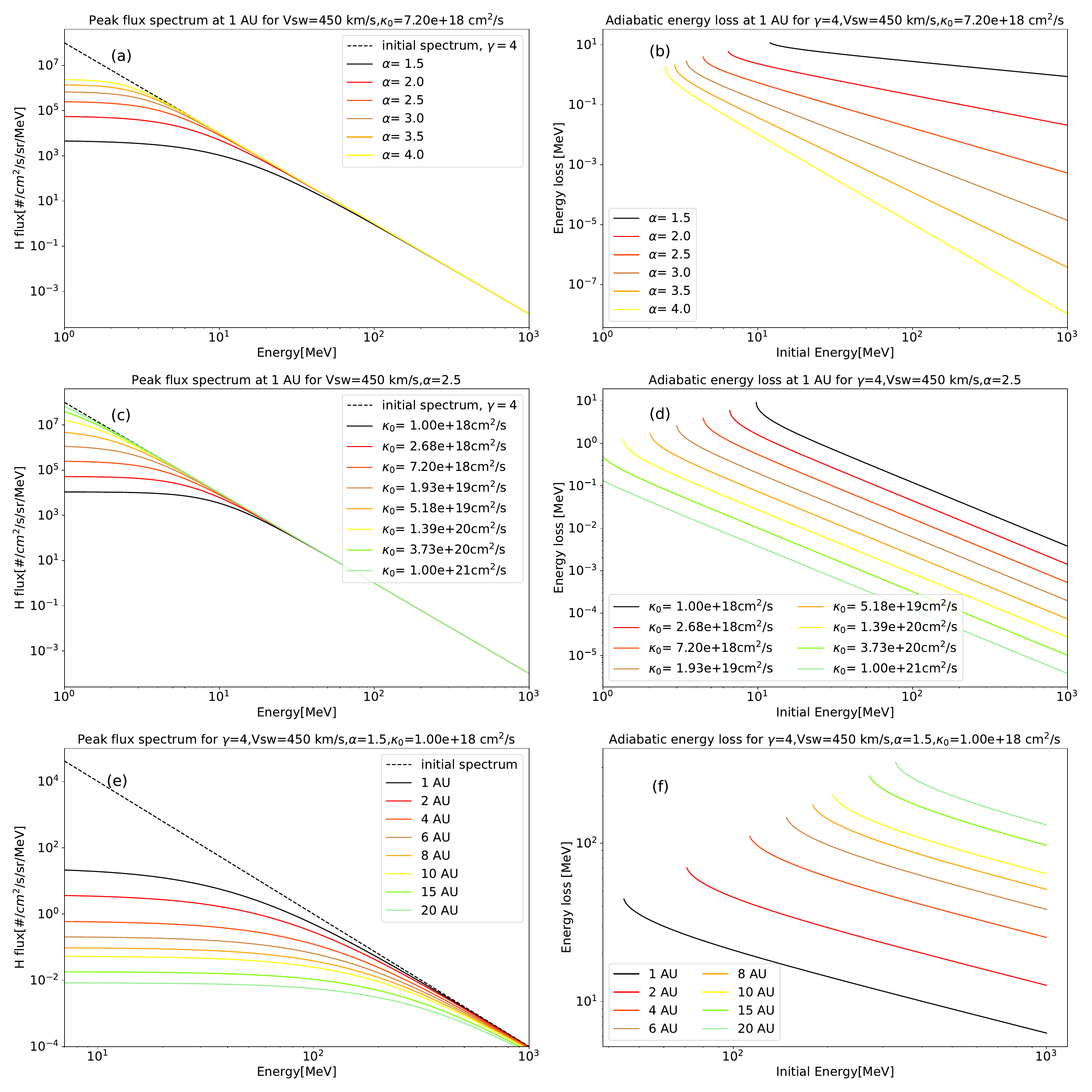} 
\caption{Impact of adiabatic energy loss on the particle energy spectrum as described by Eq. \ref{eqn:peak_spectrum}. We set an initial flux spectrum with parameters $j_0=10^8 [\mathrm{\#/cm^2/s/sr/MeV}]$, $\gamma=4$, and a constant solar wind speed of $V_{sw}$=450 [km/s]. Panel a shows the peak flux spectrum at 1 AU for different diffusion coefficient index $\alpha$. Panel b shows the adiabatic energy loss of particles with a given initial energy corresponding to Panel a. Panel c shows the peak flux spectrum at 1 AU for different diffusion normalization coefficient $\kappa_0$. Panel d shows the adiabatic energy loss of particles with a given initial energy corresponding to Panel c. Panel e shows the evolution of the peak flux spectrum at different heliocentric distance. Panel f shows the adiabatic energy loss corresponding to Panel e.}
\label{fig:peak_spectra_model}
\end{minipage}
\end{figure*}

 Figure \ref{fig:peak_spectra_model} illustrates the impact of adiabatic energy loss on the spectrum as described by the above equation. We set an initial flux spectrum with parameters $j_0=10^8 [\mathrm{\#/cm^2/s/sr/MeV}]$, $\gamma=4$, and a constant solar wind speed of $V_{sw}$=450 [km/s]. Panel a shows the peak flux spectra at 1 AU for different diffusion coefficient index $\alpha$. Panel b shows the adiabatic energy loss of particles with a given initial energy corresponding to Panel a. Panel c shows the peak flux spectra at 1 AU with different diffusion normalization coefficient $\kappa_0$. Panel d shows the adiabatic energy loss of particles with a given initial energy corresponding to Panel c. Panel e shows the evolution of the peak flux spectrum at different heliocentric distance. Panel f shows the adiabatic energy loss corresponding to Panel e. From this graph, it can be seen that smaller diffusion coefficients, namely smaller $\kappa_0$ and $\alpha$, will have a more pronounced adiabatic cooling effect on particles, especially for those with lower energy (less than tens of MeV). Additionally, as the propagation distance increases, the peak energy spectrum of SEP events will also become flatter.

\section{SEPEM dataset and methods}\label{sec:SEPEM dataset and Methods}
The Solar Energetic Particle Environment Modelling (SEPEM) project is a WWW interface to SEP data together with a range of modelling tools and functionalities intended to support space mission design\footnote{\url{http://sepem.eu/}\label{url:SEPEM}}. The system provides an implementation of several well-known modelling methodologies, built upon cleansed datasets. A large number of datasets have been combined into an SQL database for convenient access. SEPEM also affords the user increased flexibility in their analysis and enables the generation of mission-integrated fluence statistics, peak flux statistics, and other functionalities. Additionally, it integrates effect tools that calculate single-event upset rates and radiation doses for a variety of scenarios; the statistical methods can further be applied to these effect parameters.
\par In this work, we exploit the continuous and high-quality dataset and the long-term reference SEP event list to statistically study the properties of SEP spectra.

\subsection{SEPEM dataset}\label{sec:SEPEM dataset}
The SEPEM reference proton dataset was originated from multiple spacecrafts and instruments\footnote{\url{http://sepem.eu/help/data_sources.html}\label{url:SEPEM_data_source}}, such as Space Environment Monitor (SEM) and Energetic Particles Sensor (EPS) of Geostationary Operational Environmental Satellite (GOES), and Goddard Medium Energy (GME) of Interplanetary Monitoring Platform 8 (IMP 8) satellite and further processed through correction, completion, cross-calibration, and energy rebinning \citep{crosby2015sepem}.
\par We selected the most up-to-date version (3rd version) of the SEPEM reference proton dataset for this study. This dataset includes proton flux data at 1AU with a time resolution of 5 minutes, spanning from July 1, 1974 to December 31, 2017. It also consists of 14 logarithmic energy bins ranging from 5~MeV to nearly 900~MeV, with Galaxy Cosmic Ray (GCR) background subtracted. For each energy channel we use the geometric mean of upper and lower limit as the effective energy (see Table \ref{tab:SEPEM_energy_bins}).

\begin{table*}
\caption{Fourteen logarithmic energy bins of the third version of the SEPEM reference proton dataset.}
\begin{center}
\begin{tabular}{cccccc}
\toprule  
\multirow{2}*{Channel} & Effective energy& \multirow{2}*{Channel} & Effective energy &\multirow{2}*{Channel} & Effective energy \\
 &(Energy range) [MeV] &  & (Energy range) [MeV] &  & (Energy range) [MeV]\\
\midrule  
1&6.01(5.00-7.23)&6&38.03(31.62-45.73)&11&244.2(200.0-289.2)\\
2&8.70(7.23-10.46)&7&54.99(45.73-66.13)&12&347.8(289.2-418.3) \\
3&12.58(10.46-15.12)&8&79.53(66.13-95.64) &13& 503.0(418.3-604.9)\\
4&18.18(15.12-21.87)&9&115.0(95.64-138.3)&14&727.4(604.9-874.7) \\
5&26.30(21.87-31.62) &10&166.3(138.3-200.0)& &\\
\bottomrule 
\end{tabular}\label{tab:SEPEM_energy_bins}
\end{center}
\end{table*}

\par We note that the uncertainty in the flux values is not provided in the SEPEM dataset and it is nontrivial to derive it as the flux is rebinned into energy bins different from those used directly for measurements. Since the resulting uncertainties on the fit parameters do not affect our subsequent analysis and results, we assign an arbitrary uncertainty of 5\% to all energy channels to complete the fitting algorithm following previous studies \cite[e.g.,][]{dierckxsens2015}.

\subsection{SEPEM reference event list}\label{sec:SEPEM reference event list}
At present, there is no world-wide consistent definition of the start and end time for SEP events. The SEPEM project selected 7.23–10.46~MeV proton channel as the reference channel and generated a reference SEP event list\footnote{\url{http://sepem.eu/help/event_ref.html}\label{url:SEPEM_event_list}} based on the following standards \citep{Jiggens2011}: (1) Threshold for the start and end of an event: 0.01 $[\mathrm{\#/cm^2/s/sr/MeV}]$. (2) Minimum peak flux: 0.5 $[\mathrm{\#/cm^2/s/sr/MeV}]$. (3) Minimum event duration and dwell time: 24 h.  

\begin{figure*}[ht!]
\begin{minipage}[t]{1.0\linewidth}
\centering
\includegraphics[width=0.95\hsize]{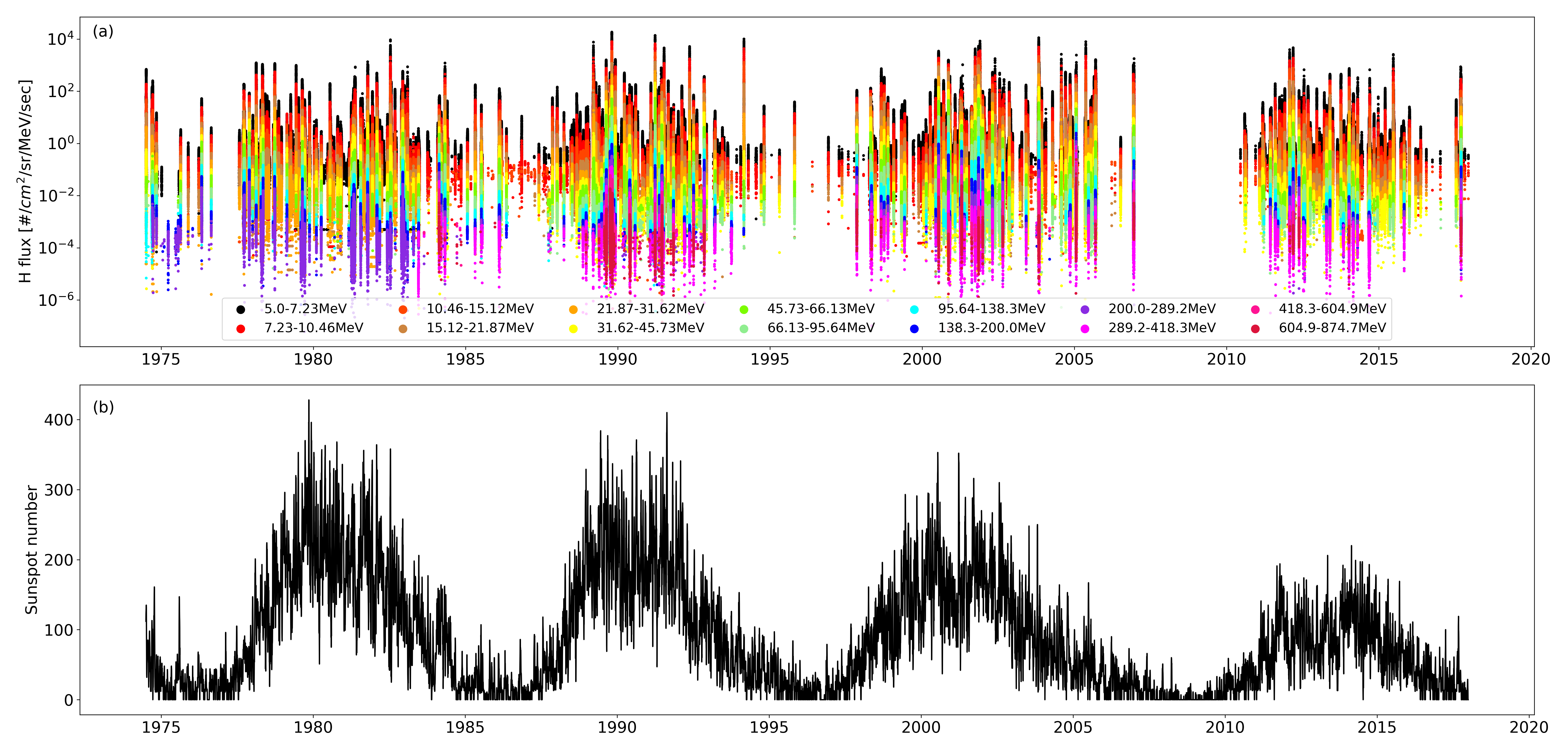} 
\caption{Occurrence of SEP events is related to the solar activity cycle. Panel a shows the entire SEPEM dataset used in this study, namely the SEP proton flux-time profile from July 1, 1974 to December 31, 2017. Each vertical colored strip represents an SEP event, with different colors indicating various energy bins. Panel b shows the corresponding daily Sunspot Number (SSN) for the same time duration. The frequency of SEP events is positively correlated with SSN indicating solar activities.}
\label{fig:SEPEM_dataset_and_SSN}
\end{minipage}
\end{figure*}
\par Fig. \ref{fig:SEPEM_dataset_and_SSN} shows the entire SEPEM dataset used in this study (panel a), along with the corresponding daily Sunspot Number (SSN) for the same time period (panel b). The SSN data are accessed from the World Data Center SILSO\footnote{Solar Influences Data Analysis Center \url{https://www.sidc.be/SILSO/datafiles}\label{url:SSN_website}}, Royal Observatory of Belgium, Brussels \citep{sidc}. As expected, more SEP events occur during periods of high solar activity. There are also quite a number of events occurring in the rising and decay phases of solar cycles. This is a well-known feature of SEP events and poses great challenges to the prediction of the occurrence and intensity of SEP events.

\par By the definition of single SEP event in the original SEPEM reference event list, there may be more than one enhancements from several events (namely, the “single” event can be further divided into several events with individual rising, peak and declining phases). This could lead to an incorrect peak flux selection. So we modified the start and end times of the original SEPEM reference event list to make both the peak energy spectrum and integrated energy spectrum related to the same SEP event. More details can be found in Sect. \ref{sec:result}, Table 3 and Table 4\footnote{Table 3 and Table 4 can be downloaded from \url{https://doi.org/10.5281/zenodo.13270536}\label{url:Table 3 and Table 4}}.

\subsection{Spectra Fitting Methods}\label{sec:Methods}
Considering the forms of Eqs. \ref{eqn:Band_function} and \ref{eqn:peak_spectrum}, we utilize the nonlinear least squares method to obtain the optimal fitting parameters by minimizing the sum of squares error function  $S(\boldsymbol{\theta})$ with the following definition \citep{Vugrin2007}:
\begin{align}
S(\boldsymbol{\theta})=\sum_{i=1}^{n}\left[f\left(\boldsymbol{\theta} ; \mathbf{x}_{\mathbf{i}}\right)-y_{i}\right]^{2}=\sum_{i=1}^{n}\left[r_{i}(\boldsymbol{\theta})\right]^{2}, \label{eqn:least_square}
\end{align}
where $f\left(\boldsymbol{\theta} ; \mathbf{x}_{\mathbf{i}}\right)$ is the nonlinear model, $\boldsymbol{\theta}$ is a vector of all parameters, and $r_{i}(\boldsymbol{\theta})$ is residual.
\par
\begin{figure*}[ht!]
\begin{minipage}[t]{1.0\linewidth}
\centering
\includegraphics[width=0.95\hsize]{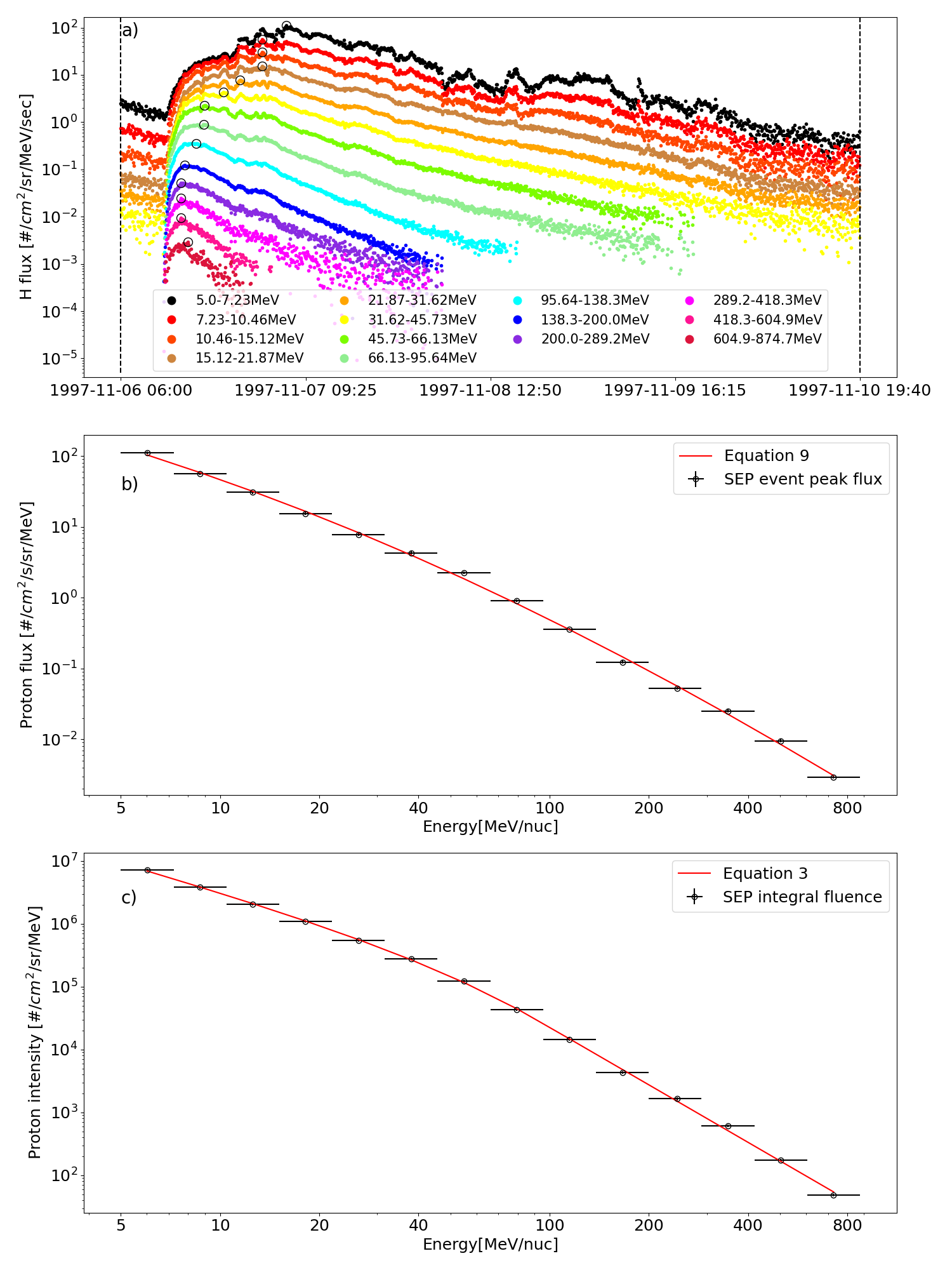} 
\caption{Example of spectrum fitting. Panel a shows the flux time profile of a SEP event during November 1997. The black circles represent the peak flux of each energy bin. The vertical dashed lines indicated the start and end time for the integrated fluence spectrum. Panel b and c show the peak flux spectrum and the integrated fluence spectrum fitting, respectively.}
\label{fig:example}
\end{minipage}
\end{figure*}

 Figure \ref{fig:example} shows an example of spectrum fitting. Panel a shows the flux time profile of a SEP event during November 1997. The black circles represent the peak flux of each energy bin. Panel b shows the peak flux spectrum fitting. Panel c shows the integrated fluence spectrum fitting throughout the event.

\par We note that there are still tens of events whose spectra can be fitted by neither Eq. \ref{eqn:Band_function} nor Eq. \ref{eqn:peak_spectrum}. There are multiple reasons for this, such as significant impact of the previous SEP event, influence of magnetic field structures on the SEP flux evolution, or other obvious instrumental or systematic errors. These events are excluded from the table and the following study.

\section{Results and discussions}\label{sec:result}
\par Fitting parameters for the peak flux spectra following Eq. \ref{eqn:peak_spectrum} are listed in Table 3.  As mentioned in Sect. \ref{sec:SEPEM dataset}, the uncertainties are calculated assuming a 5\%  arbitrary dataset uncertainty. $\kappa_0$\_450 is calculated using a constant solar wind speed of 450 [km/s], while $\kappa_0$ is calculated using the mean solar wind speed in the 6 days after the events onset. The corresponding solar wind speed values and standard deviations are listed in columns "Vsw" and "Vsw\_err". The solar wind speed data is obtained from GSFC/SPDF OMNIWeb interface\footnote{\url{https://omniweb.gsfc.nasa.gov}\label{url:omni_web}}.  We also obtained corresponding X-ray flare class for each SEP event from NOAA space environment services center\footnote{\url{https://umbra.nascom.nasa.gov/SEP/}\label{url:NOAA_sep}} for further analysis in Sect. \ref{sec:Relationship with x-ray flare class}. 
\par Fitting parameters for the time-integrated fluence spectra, namely Eq. \ref{eqn:Band_function}, are listed in Table 4 where $E_{break}=E_{0}\left(\gamma_{1}-\gamma_{2}\right) $ is the transition energy of the Band function. The GLE event list is obtained from the Neutron Monitor GLE database\footnote{\url{https://gle.oulu.fi/\#/}\label{url:GLE_database}}.
\par The calculation of the chi-square value for energy spectrum fitting is as follows:
\begin{align}
\chi^2=\frac{1}{N} \sum_{k=1}^{N} \frac{\left[j_{\text {Fit}}\left(E_{k}\right)-j_{\text {Meas}}\left(E_{k}\right)\right]^{2}}{\sigma_{k}^{2}}, \label{eqn:chi_square}
\end{align}
in which N is the number of energy bins, $j_{\text {Fit}}\left(E_{k}\right)$ and $j_{\text {Meas}}\left(E_{k}\right)$ are the fitted and measured particle flux at the kinetic energy of $E_{k}$, respectively. $\sigma_{k}=0.05j_{\text {Meas}}\left(E_{k}\right)$ represents the aforementioned uncertainty on the data in the $\text{k}^{th}$ bin. 
\par As we can see from Table 3 and Table 4, there are a significant number of events for which the value of $\chi^2$ is relatively large. The reasons for large $\chi^2$ value may be divided into two categories: One comes from the data processing algorithms. Data of SEP flux in the four highest energy channel of Table \ref{tab:SEPEM_energy_bins} can be significantly influenced by the GCR background subtract algorithms, which makes the fitting of this part difficult. The other reason is the influence of local acceleration and propagation effects which are not considered in the model used here.

\subsection{Distribution of spectral parameters}\label{sec:Distribution of parameters}
\par We analyze the peak flux spectra of 103 events and the integral fluence spectra of 164 events in total. The other 160(99) events listed in the reference event table are not involved in the following analysis as mentioned in Sect. \ref{sec:Methods}.

\begin{figure*}
\begin{minipage}[t]{1.0\linewidth}
\centering
\includegraphics[width=0.95\hsize]{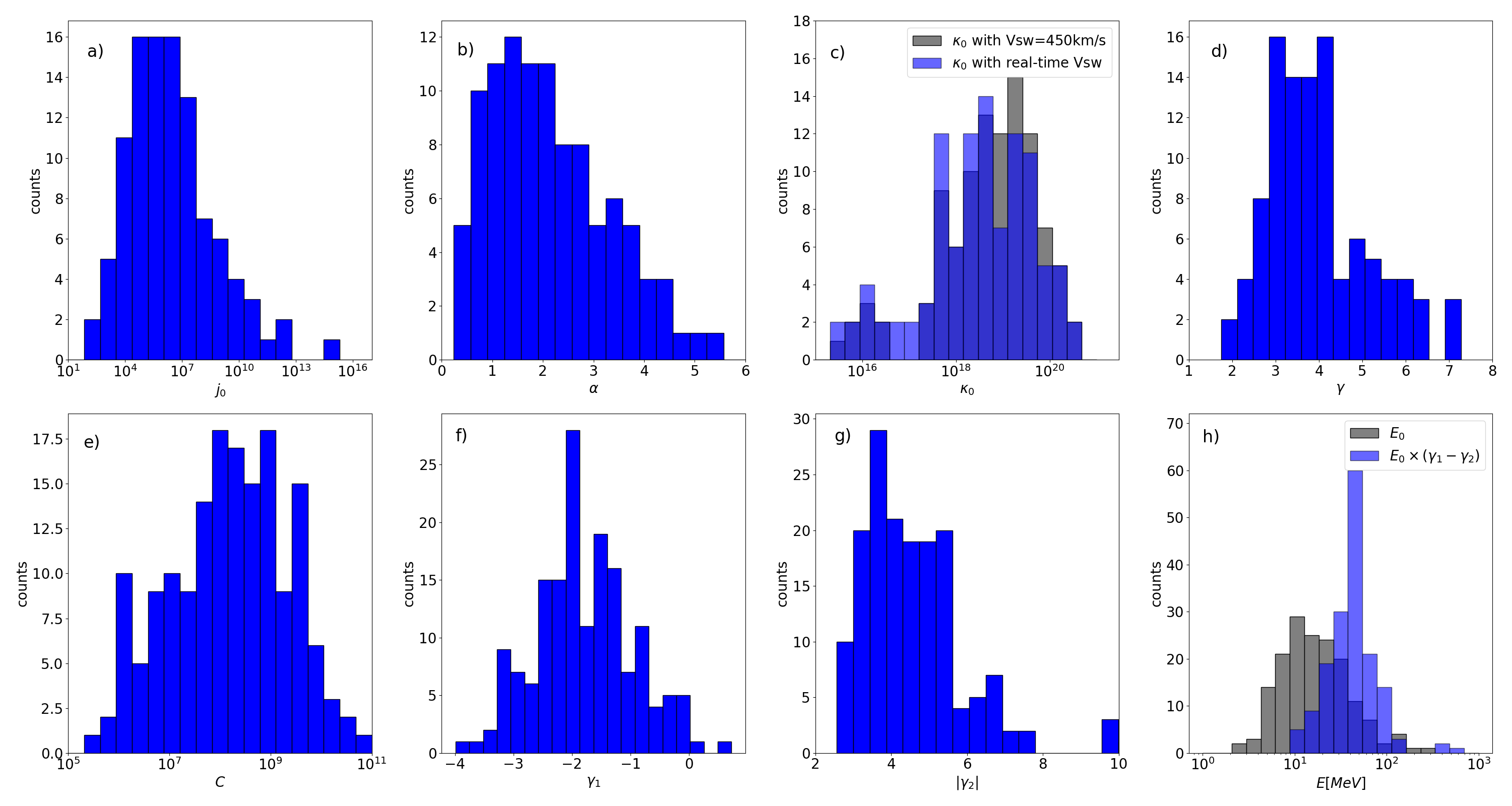} 
\caption{Histograms of the fitting parameters shown in Table 3 and Table 4. The results of the Anderson–Darling test are shown in Table \ref{tab:A-D test}, indicating that logC, $\gamma_1$, and log$E_0$ follow a normal distribution.}
\label{fig:results1}
\end{minipage}
\end{figure*}
\par The histograms in Fig. \ref{fig:results1} show the distribution of fitting parameters listed in Table 3 and Table 4. Panel c illustrates the distribution of $\kappa_0$\_450 obtained through the utilization of a constant solar wind speed of 450 [km/s], and $\kappa_0$ derived from real-time solar wind speed observations near Earth. Our subsequent analysis relies on the $\kappa_0$ derived from real-time solar wind speed. The histograms show that the utilization of a constant solar wind speed does not appear to have a significant impact on the distribution of this parameter, but it may differ significantly for individual events as shown in Table 3.
\par To better quantify their distribution feature, we further investigated the distribution of fitting parameters using the Anderson–Darling test \citep{anderson1954test} with the results shown in Table \ref{tab:A-D test}. It can be seen that the statistical values of parameters logC, $\gamma_1$, and log$E_0$ are less than the critical value for a significance level of 0.05. This indicates that they mostly follow a normal distribution. More discussions will be given in Sect. \ref{sec:summary}

\begin{table*}
\caption{Anderson–Darling test  (significance level equals to 0.05) for fitting parameters.}
\begin{center}
\begin{tabular}{cccccc}
\toprule  
Parameter &statistic&critical value & Parameter &statistic&critical value\\
\midrule  
log$j_0$&1.23&0.76&logC&0.49&0.77\\
$\alpha$&1.71&0.76&$\gamma_1$&0.41&0.77\\
log$\kappa_0$&1.01&0.76&$\gamma_2$&4.94&0.77\\
$\gamma$&1.83&0.76&log$E_0$&0.68&0.77\\
\bottomrule 
\end{tabular}\label{tab:A-D test}
\end{center}
\end{table*}

\subsection{Relationships between integral fluence spectra and peak flux spectra}\label{sec:Connections of two spectra}
\par We now try to find out whether the peak flux spectrum and integral fluence spectrum of the same SEP event have a certain kind of correlation. Fig. \ref{fig:results2} demonstrates the relationships among 8 fitting parameters. For each SEP event indicated by a dot or circle, the four parameters of the peak flux spectrum  (Eq. \ref{eqn:peak_spectrum}) are plotted on the x-axes, while the four parameters of the integral fluence spectrum (Eq. \ref{eqn:Band_function}) are plotted on the y-axes. Solid dots are fitted parameters of GLE events, while parameters of other SEP events are represented by hollow circles. The red line represents the linear regression curve of two parameters for both GLE events and non-GLE events, and the corresponding linear regression function is shown in the legend. The shaded area represents the confidence interval for one standard deviation. From the graph, we can summarize some interesting relationships between the two sets of parameters:

\begin{figure*}[ht!]
\begin{minipage}[t]{1.0\linewidth}
\centering
\includegraphics[width=0.95\hsize]{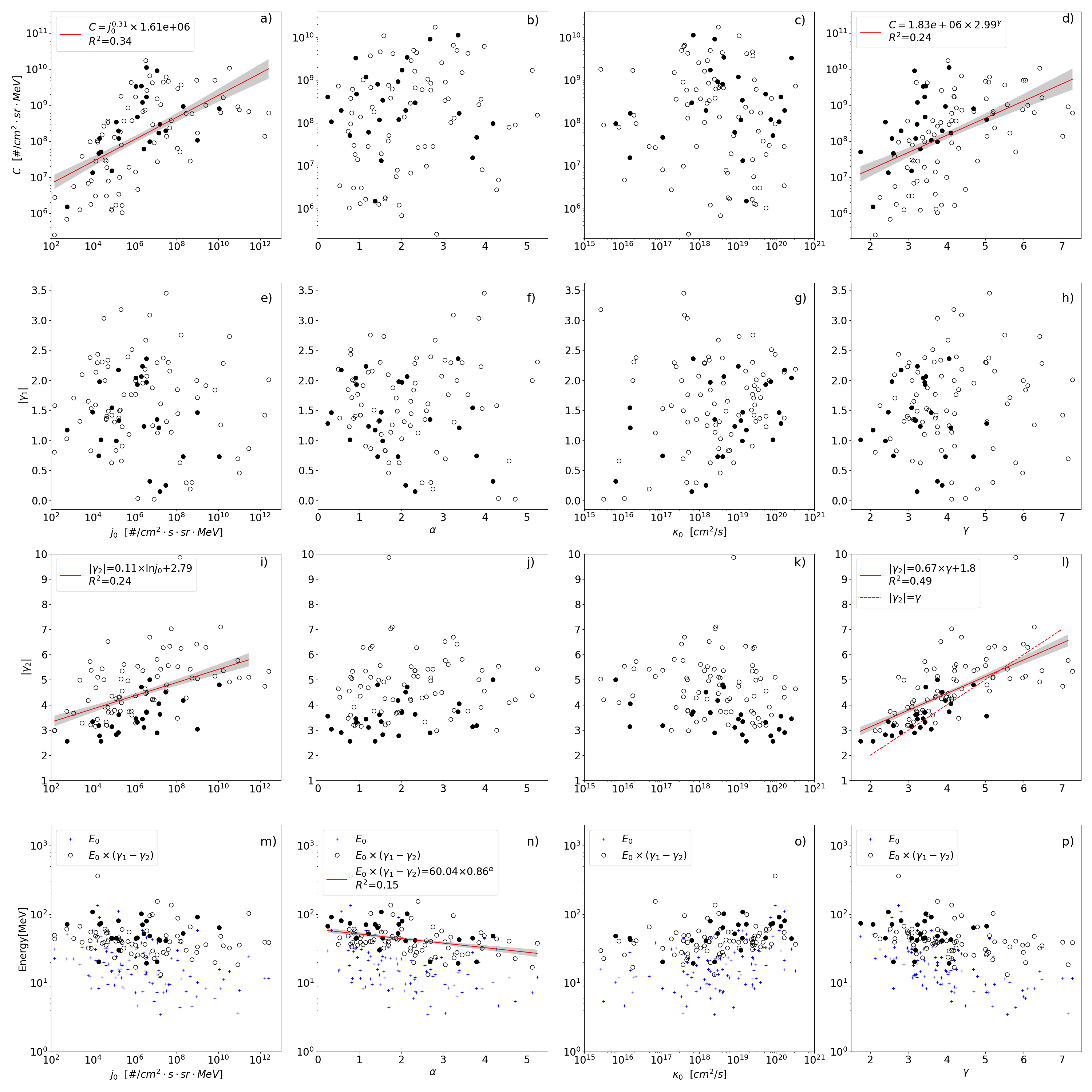} 
\caption{Relationships among 8 fitting parameters. For each event, the four parameters of the peak flux spectrum  (Eq. \ref{eqn:peak_spectrum}) are plotted on the x-axes, while the four parameters of the integral fluence spectrum (Eq. \ref{eqn:Band_function}) are plotted on the y-axes. Solid dots are parameters of GLE events, while parameters of other SEP events are represented by hollow circles. In each panel where the $R^2$ is larger than 0.1, the red line represents the linear regression curve between two parameters for all events, and the corresponding linear regression function is shown in the legend. The shaded area represents the confidence interval for one standard deviation. The red dashed line in panel l represents the function of $|\gamma_2|=\gamma$.}
\label{fig:results2}
\end{minipage}
\end{figure*}

\begin{enumerate}
    \item\label{res:relationships1}In panel a, the two normalization coefficients, C and $j_0$, have a clear positive correlation. This is reasonable: if more SEPs are released at the source region, it is expected that a greater number of SEPs will be detected at Earth. Once SEPs are released from the acceleration source, they will propagate in different directions. However, we only have data for the Earth's orbit from the SEPEM dataset, which could result in the dispersion of this positive correlation.
    \item\label{res:relationships2_1}In panel d, the normalization coefficients C and SEPs' source spectrum index $\gamma$ also have a positive correlation indicating that events with larger fluence at lower energy range more likely result from a softer SEP spectrum. There may be two reasons for this. One reason is that the SEPEM reference event list has a strict definition of SEP events  (see Sect. \ref{sec:SEPEM reference event list}) with the original intention to exclude particle enhancement events of non-solar origin from the list, such as those accelerated by Stream Interaction Regions. This, however, has resulted in some weaker SEP events also being excluded from the list. These events have a smaller normalization coefficients C and a larger spectrum index $|\gamma|$ so that they do not have enough particles within the energy range of 7.23-10.46 MeV to define an SEP event in the SEPEM list. This can explain why there is no distribution of data points in the lower right part of this panel.
    \item \label{res:relationships2_2} Panel d also shows that there lack events with both very large intensities and small $\gamma$ values. An early study by \cite{fichtel1967energetic} estimated that the total energy of particles in a large SEP event was about $10^{30}$[ergs], roughly 1 percent of total energy of a flare. \cite{emslie2012global} evaluated the energy budget of 38 large solar eruptive events between 2002 and 2006, and found that the SEP energy was between $\sim 0.1\%$ and $3.5\%$ of the total eruptive energy. For instance, they estimated that SEPs in GLE event Nr. 65 carried the highest energy of $4.3\times10^{31}$[ergs], which was about 1.5\% of the estimated free magnetic energy ($\sim 3\times10^{33}$[ergs]) of the X17 flare. 
    The SEP spectrum is constrained by the total energy of accelerated particles: $\int Ejd{E}= \frac{j_0}{\gamma-2}(E_{min}^{2-\gamma}-E_{max}^{2-\gamma})$. For a given energy limit, the larger the intensity scaling factor $j_0$ (flux at 1 MeV), the steeper the spectrum (with a larger $|\gamma|$).
     \item\label{res:relationships3}In panel l, we can observe a strong positive correlation between the spectral indices of the high-energy portion of the integral spectra $|\gamma_2|$ and the initial spectra $\gamma$ and we fit them as $|\gamma_2|=0.67\times\gamma+1.8$. 
     The results of linear regression is not far from the function of $|\gamma_2|=\gamma$, which is represented by the red dashed line. 
     This reason that $|\gamma_2|$ differs from $\gamma$ for single events may be a result of the combined effects of adiabatic energy loss and multiple crossing effects related to particle scattering \citep{chollet2010effects}. The former tends to result in a softer spectrum ($|\gamma_2|>\gamma$) as higher-energy particles lose their energy to become lower-energy ones, while the latter modifies a spectrum to become harder ($|\gamma_2|<\gamma$) as high-energy particles are more likely to be observed multiple times at a given location. 
     \item Alternatively, the correlation between $\gamma$ and $\gamma_1$ as shown in panel h is less obvious. This is because lower energy particles experience more transport effects given their smaller mean free path length \citep{lario2007}. These effects include the aforementioned adiabatic energy loss and multiple crossing as well as cross-field diffusion and pitch angle scattering.  
     \item\label{res:relationships4}Combining panel l and panel i, it can be observed that when there is an enhancement of particle acceleration in the SEP source region (namely, larger $j_0$), the SEP source spectrum tends to become softer (namely, larger $\gamma$ and $\left|\gamma_2 \right|$). This suggests that there might be some energy constraints in the process of SEP acceleration as mentioned in Item \ref{res:relationships2_2}, or that enhanced acceleration tends to prioritize the energization of low-energy particles as it takes longer time to accelerate particles to higher energies according to the diffusive shock acceleration mechanism \citep{decker1986numerical}.
     \item\label{res:relationships5}Four panels at the bottom show the relationship between peak spectral parameters and the transition energy of the Band function: $E_{break}=\left(\gamma_{1}-\gamma_{2}\right) E_{0}$. A larger diffusion coefficient power-law index $\alpha$ or a larger SEP event source energy spectrum index $\gamma$ tends to result in a smaller transition energy. 
     This indicates that the transition energy may be a result of the propagation process. Considering adiabatic cooling process which depends on $\kappa$ and $\alpha$, we have shown that the peak energy flux of the low-energy region is more significantly reduced with smaller $\alpha$, as unveiled in panel a of Fig. \ref{fig:peak_spectra_model}, which results in a larger transition energy in the Band function.
    \item\label{res:relationships6} GLE events are registered when sufficient number of high-energy particles can overcome Earth's magnetic shielding and generate secondary particles in Earth's atmosphere to trigger ground neutron detectors. Thus, by definition, GLE spectra contain an enhanced high-energy component. This is shown in our plots: As marked by solid dots in each panel, GLE events typically have larger overall fluence normalization coefficients, C, harder SEP source spectra (smaller $\gamma$), as well as harder integral fluence spectra of the high-energy part (smaller $\left|\gamma_2\right|$) and higher transition energies of the Band function.    
\end{enumerate}

\subsection{Spectral evolution over the solar cycle}\label{sec:Distribution on time}

\begin{figure*}
\begin{minipage}[t]{1.0\linewidth}
\centering
\includegraphics[width=0.95\hsize]{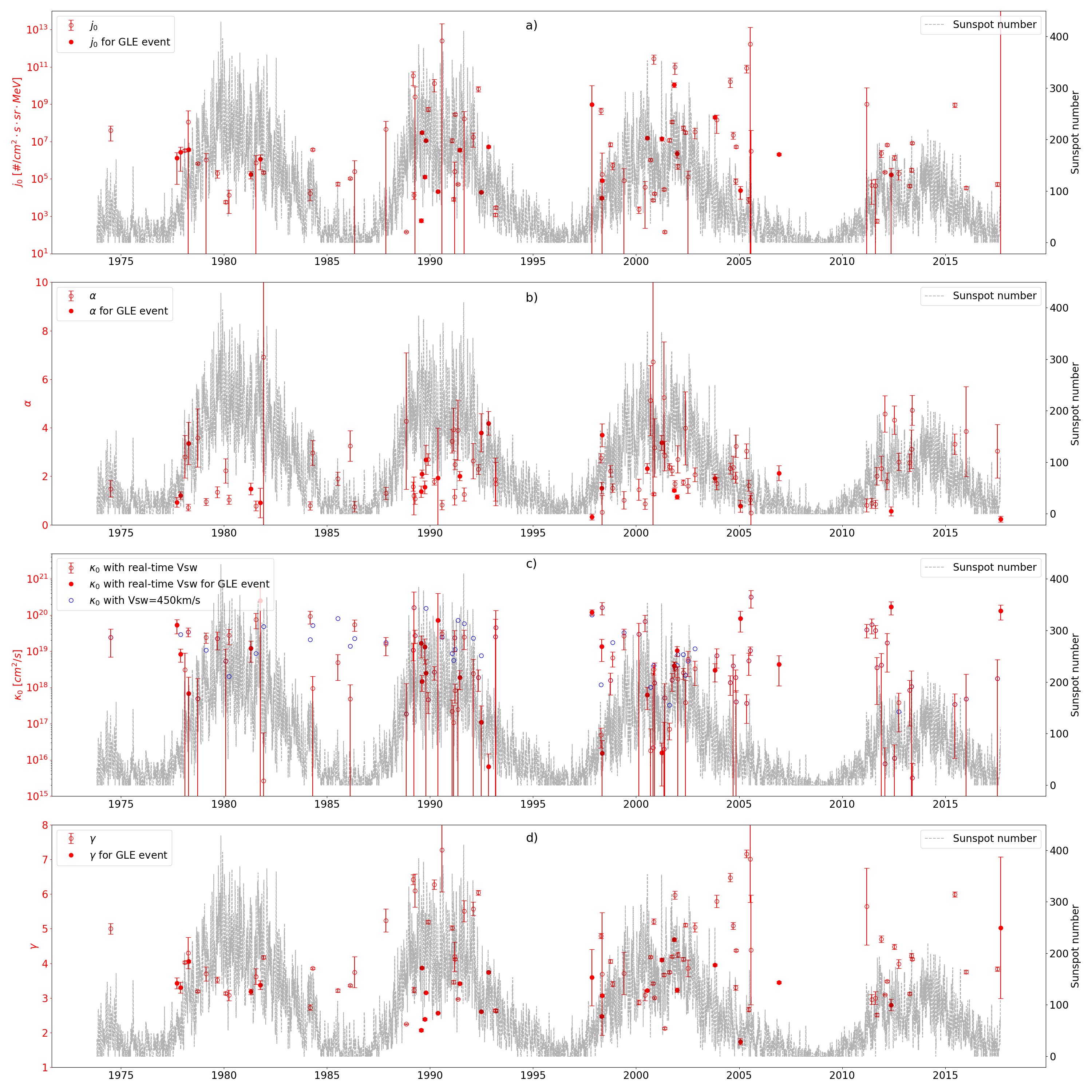} 
\caption{Solar-cycle evolution of four parameters in the peak flux spectrum as shown by the left y-axes, namely, $j_0$, $\alpha$, $\kappa_0$, $\gamma$ in Eq. \ref{eqn:peak_spectrum} for a, b, c, d panels respectively. The uncertainty of each parameter results from assuming a 5\% uncertainty in the SEPEM dataset as mentioned in Sect. \ref{sec:SEPEM dataset}. Solid dots are parameters of GLE events. The black lines in the background represent the daily sunspot number as scaled by the right y-axes. }
\label{fig:results3}
\end{minipage}
\end{figure*}

\begin{figure*}[ht!]
\begin{minipage}[t]{1.0\linewidth}
\centering
\includegraphics[width=0.95\hsize]{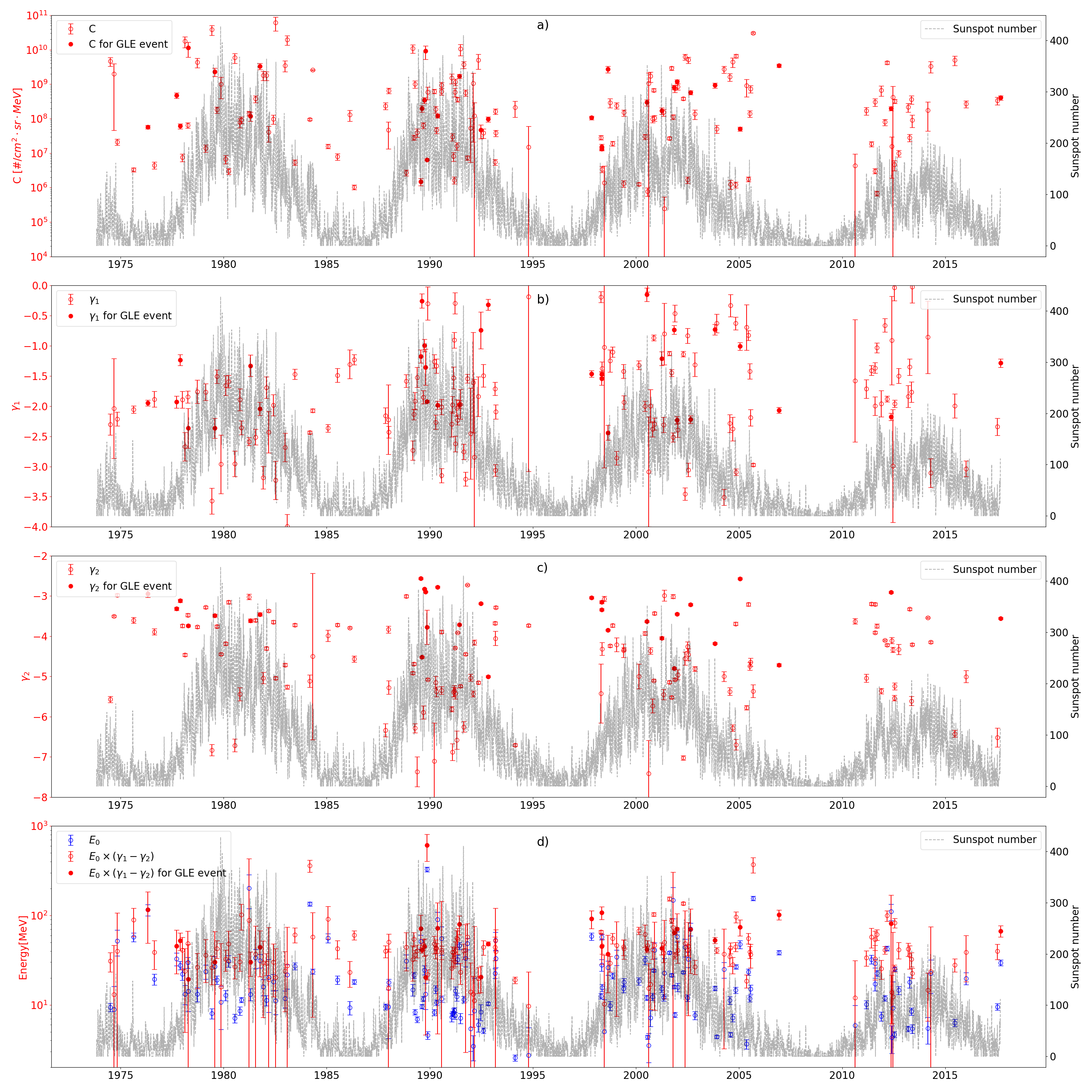} 
\caption{Solar-cycle evolution of four parameters in the integral fluence spectrum in Eq. \ref{eqn:Band_function} as shown by the left y-axes. The uncertainty of each parameter results from assuming a 5\% uncertainty in the SEPEM dataset as mentioned in Sect. \ref{sec:SEPEM dataset}. Solid dots are parameters of GLE events. The black lines in the background represent the daily sunspot number as scaled by the right y-axes. }
\label{fig:results4}
\end{minipage}
\end{figure*}

\begin{figure*}[ht!]
\begin{minipage}[t]{1.0\linewidth}
\centering
\includegraphics[width=0.95\hsize]{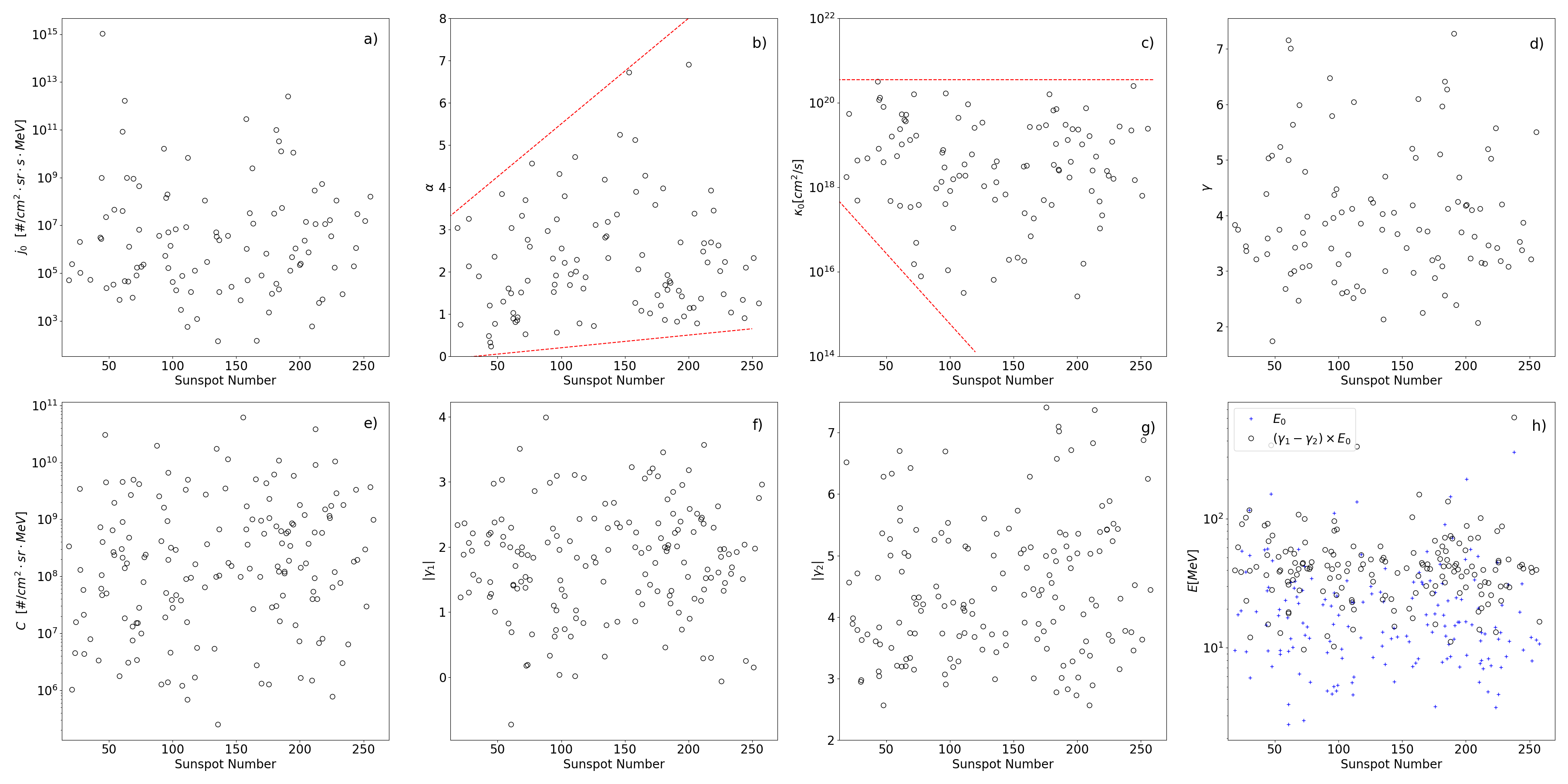} 
\caption{Distribution of the energy spectrum parameters versus the sunspot number. We use the monthly-averaged SSN centered at each SEP event's start time. The red dashed lines illustrate the distribution boundaries of $\alpha$ and $\kappa_0$ with respect to sunspot number.}
\label{fig:results5}
\end{minipage}
\end{figure*}

\begin{figure*}[ht!]
\begin{minipage}[t]{1.0\linewidth}
\centering
\includegraphics[width=0.95\hsize]{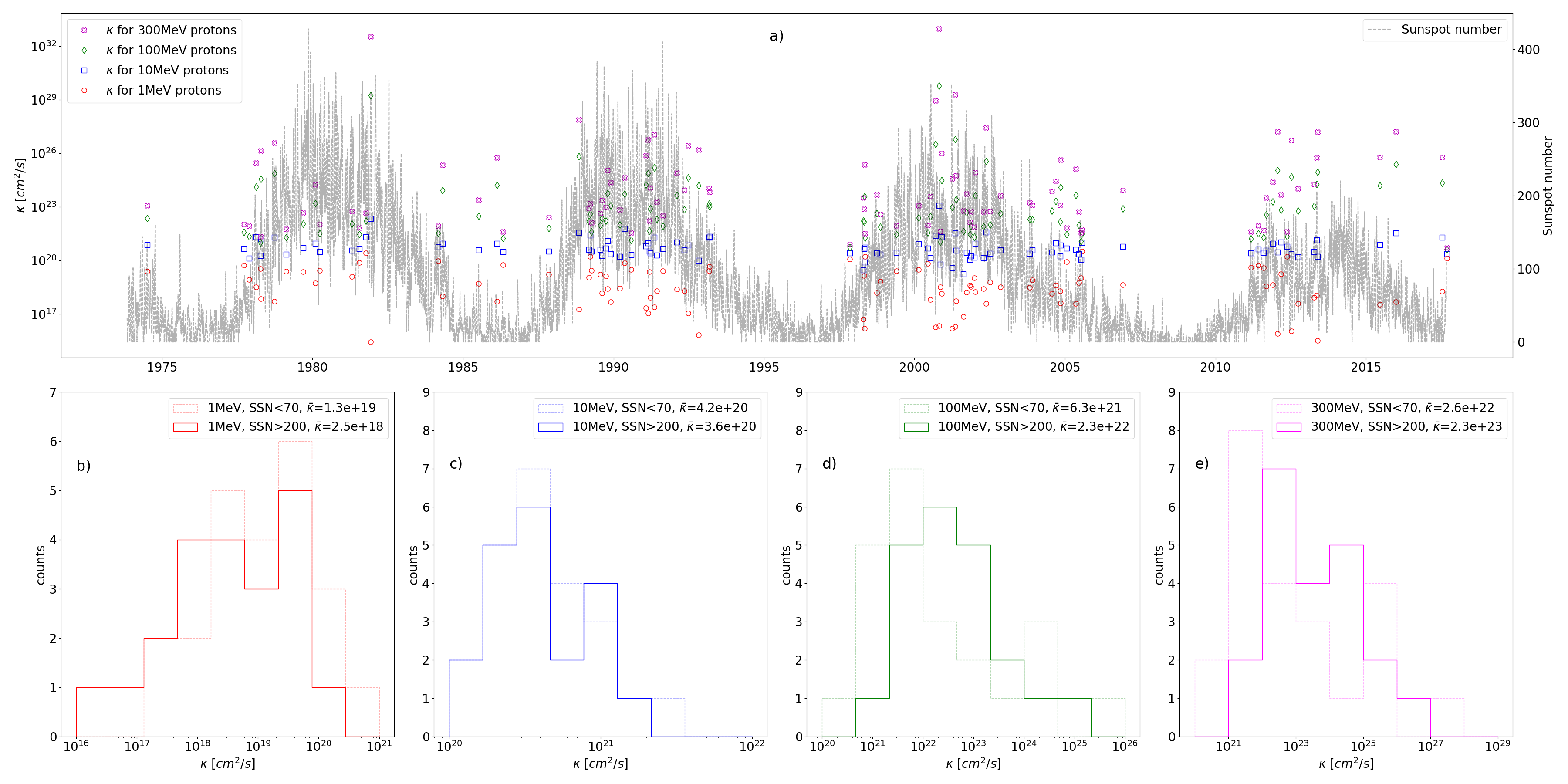} 
\caption{Solar-cycle dependent diffusion coefficient of different energies. Panel a shows the solar cycle evolution of the diffusion coefficient (left y-axis), which is defined by Eq. \ref{eqn:diffusion coefficient}, for four different energies of protons (left legends). The gray lines in the background represent the daily sunspot number as scaled by the right y-axis. Panels b, c, d, and e show the diffusion coefficient distribution of 10, 50, 100, and 300 MeV protons respectively. The distributions for low solar activity are represented by the dashed line, and for high solar activity are represented by the solid line. The median value of the diffusion coefficient $\bar{\kappa}$ for each distribution is also displayed in the legends.}
\label{fig:results5_2}
\end{minipage}
\end{figure*}

\par It has been almost 200 years since the discovery of the $\sim$11-year solar activity cycle\citep[e.g.,][]{hathaway2015}. As shown in Fig. \ref{fig:SEPEM_dataset_and_SSN}, the occurrence of SEP events is closely related to the solar activity cycle. Due to the extensive coverage of the SEP event list studied in this work, which spans almost four solar cycles, it is possible to conduct statistical research on the energy spectra of SEP events during various solar activity phases.
\par Figure \ref{fig:results3} demonstrates the time distribution of four parameters in the peak flux spectrum  (Eq. \ref{eqn:peak_spectrum}). Figure \ref{fig:results4} demonstrates the time distribution of four parameters in the integral fluence spectrum (Eq. \ref{eqn:Band_function}). Solid dots are parameters of GLE events. The black lines in the background represent the daily sunspot number. 
\par In Fig. \ref{fig:results5}, we examine the relationship between the energy spectral parameters and the sunspot number. Since the sunspot number is only a proxy for the solar activity and its daily values are highly fluctuating, we use the monthly-averaged SSN centered at each SEP event's start time, and we do not label the GLE event separately. The red dashed lines illustrate the distribution boundaries of $\alpha$ and $\kappa_0$ with respect to sunspot number. 
\par In Fig. \ref{fig:results5_2}, we further examine the diffusion coefficient evolution over the solar cycle. Panel a shows the results of the diffusion coefficient $\kappa=\kappa_{0}E^{\alpha}$ calculated for protons of 1, 10, 100, and 300 MeV. We consider monthly-averaged SSN less than 70 as low solar activity periods and those greater than 200 as high solar activity periods. The distributions of $\kappa$ of protons with different energies for the above two solar activities are compared in panels b, c, d and e.
\par We can see from Figures 6-9 that:
\begin{enumerate}
    \item\label{res:solarcycle1} The diffusion effect of SEPs is solar-cycle dependent. 
    In Fig. \ref{fig:results3}(b),(c) and Fig. \ref{fig:results5}(b),(c), we can observe a weak solar cycle evolution of the fitted parameters $\kappa_0$ and $\alpha$ which define the diffusion coefficient $\kappa$. In Fig. \ref{fig:results5_2}, we can better see that during higher solar activity periods, both $\alpha$ and $\kappa_0$ show a wider distribution: $\alpha$ can reach a larger magnitude than during solar maximum periods, while $\kappa_0$ can reach smaller values compared to solar quiet periods. This suggests that the difference in the diffusion behavior of SEPs with different energies is larger during solar maximum periods.
    \item Fig. \ref{fig:results5_2} further illustrates the cycle-dependent diffusion coefficient of different energies (1, 10, 100 and 300 MeV protons in panels b,c,d,e, respectively). Both distribution and median value of diffusion coefficient indicate that low-energy particles (1 MeV, panel b) experience enhanced diffusion (smaller $\kappa$) during higher solar activities; high-energy particles (100 and 300 MeV, panels d and e), however, experience reduced diffusion during higher solar activities. This result agrees with the modeled results of the propagation of cosmic rays, see Fig. 10 of \cite{fiandrini2021numerical} and Fig.4 of \cite{song2021numerical}. The quasi-linear theory (QLT) predicts that the two parameters $\alpha$ and $\kappa_0$ are related to the power spectrum of the interplanetary magnetic field (IMF) turbulence. The power spectral index of the IMF turbulent level varies with different levels of solar activity, leading to changes in the scattering behavior of charged particles with different energies \citep{fiandrini2021numerical}. 
    \item\label{res:solarcycle3}Other parameters, including $j_0$, $\gamma$, $C$, $\gamma_1$, $\gamma_2$ and the break energy, do not show significant changes with varying sunspot numbers. This suggests that despite of the reduced occurring frequency, the SEP events during solar minimum are not necessarily weaker than those during solar maximum. So it is equally important to predict SEP events throughout different solar activity cycles. 
\end{enumerate}

\subsection{Relationship with X-ray flare class}\label{sec:Relationship with x-ray flare class}

\begin{figure*}[ht!]
\begin{minipage}[t]{1.0\linewidth}
\centering
\includegraphics[width=0.95\hsize]{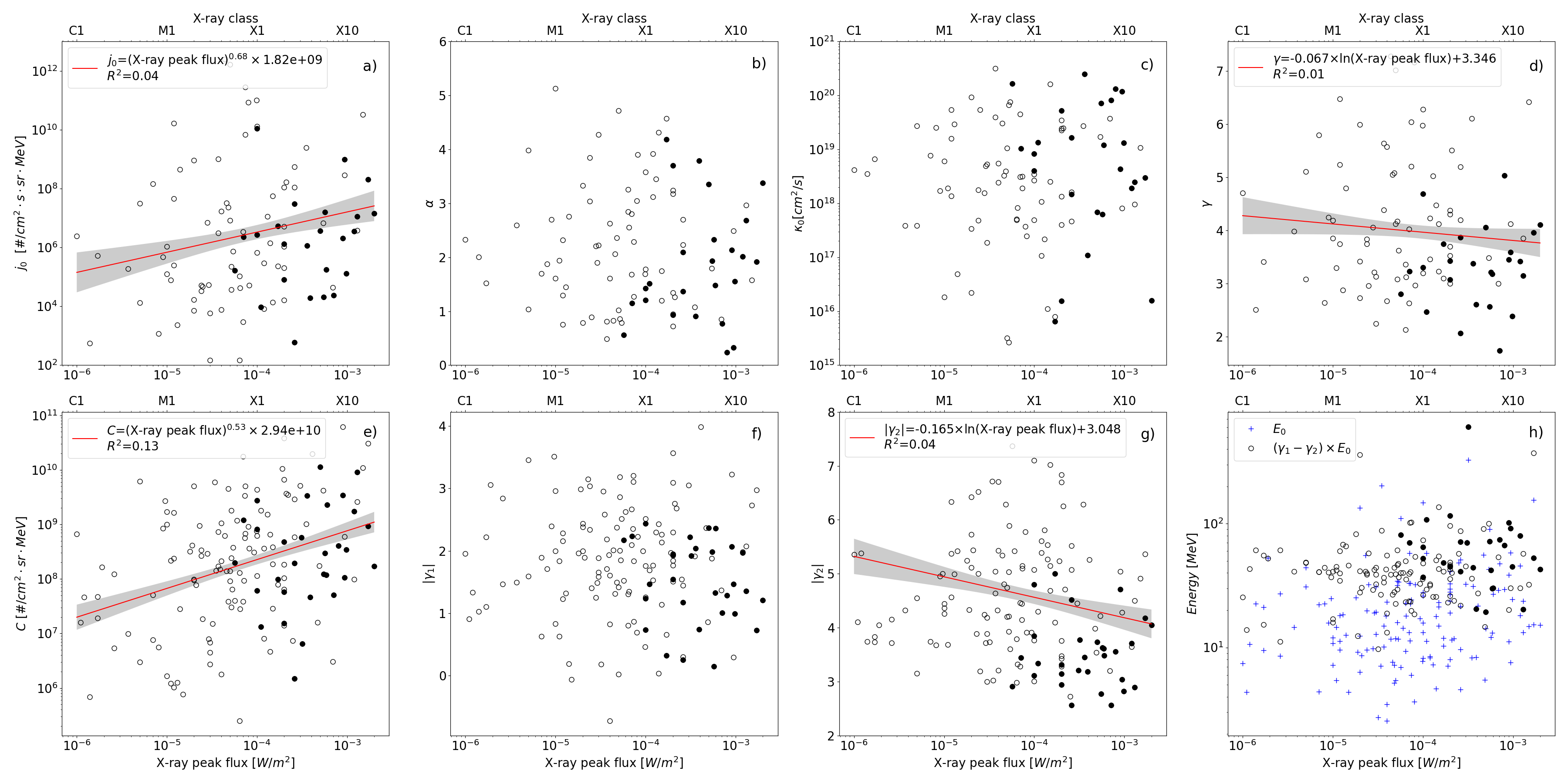} 
\caption{Relationship between the energy spectrum parameters of the SEP event and the peak X-ray flux of the corresponding flare. Solid dots are parameters of GLE events, while parameters of other SEP events are represented by hollow circles. The red line represents the linear regression curve, and the corresponding linear regression equation is shown in the legend. The shaded area represents the confidence interval for one standard deviation.}
\label{fig:results6}
\end{minipage}
\end{figure*}

\par At last, we examine the relationship between the energy spectrum parameters of the SEP event and the peak X-ray flux\footnote{Obtained from \url{https://www.swpc.noaa.gov/}\label{url:flare_xray_flux}} of the corresponding flare. As shown in Fig. \ref{fig:results6}, the following conclusions can be drawn:
\begin{enumerate}
    \item\label{res:x_ray_flux1} There is a positive correlation between the normalization coefficient of energy spectrum and the peak X-ray flux (see Fig \ref{fig:results6} (a) and (e)). This means that larger solar flares are usually accompanied by a greater flux of SEPs. This resuls agrees with previous studies \citep{belov2007peak,Kahler2007}. 
    \item\label{res:x_ray_flux2} The initial energy spectrum of SEPs, indicated by $\gamma$, and the spectrum of high-energy SEPs, indicated by $\gamma_2$, are generally harder (see Fig \ref{fig:results6} (d) and (h)) for larger solar flares. This suggests that stronger solar flares can produce more high-energy SEPs.    
    \item\label{res:x_ray_flux3} The black solid dots representing the GLE event are distributed on the right side of all panels. This is an obvious result as these events usually correspond to solar flares of higher levels. 
    \item \label{res:x_ray_flux4} As the peak X-ray flux increases, the other four parameters, $\alpha$, $\kappa_0$, $\gamma_1$, and $E_0(\gamma_1-\gamma_2)$ do not show a significant trend of change (see Fig \ref{fig:results6} (b), (c), (f), and (h)). This may be because these four parameters are mainly related to the propagation effect of SEPs in space. Even though a strong solar flare event and associated CME could sometimes alter the characteristics of solar wind plasma in interplanetary space, these changes are typically the outcome of the propagating SEPs and can be observed only later after the SEP event \citep{gopalswamy2004intensity}.
\end{enumerate}

\section{Summary and discussion}\label{sec:summary}
\par Considering the effects of diffusion-dependent adiabatic cooling on SEPs propagation, we propose a modified power-law function (Eq. \ref{eqn:peak_spectrum}) for the peak flux spectrum of SEP events. For the event-integrated spectrum, we utilize the widely-accepted Band function (Eq. \ref{eqn:Band_function}) to extract information about particle distribution. Further employing the continuous and high-quality SEPEM dataset, we are able to study SEP events covering a period of 43 years for about 4 solar cycles from 1974 to 2018. We further study the peak spectrum and event-integrated spectrum for each event and fit them with the aforementioned functions to derive the fitting parameters which indicate the physical properties as well as the acceleration and transport mechanisms of SEPs. 

\par We found that the propagation of SEPs in the inner solar system is slightly modulated by the level of solar activity, which is similar to the previous numerical research results on the propagation of cosmic rays (see Sect. \ref{sec:Distribution on time}). This helps us gain a better understanding of the propagation mechanism of charged particles in the interplanetary magnetic field. 

\par In order to further investigate the acceleration mechanism of SEPs, we examine the relationship between the energy spectrum parameters of the SEP event and the peak X-ray flux of the corresponding flare. As shown in Sect. \ref{sec:Relationship with x-ray flare class}, we found that larger SEP events, such as GLE events characterized by higher particle flux $j_0$ and harder energy spectra (smaller $\gamma$), correspond to flares of a higher-level. The power-law relationship in Fig. \ref{fig:results6} panel a, is similar to that in other studies  \cite[e.g.,][]{belov2007peak,Kahler2007} which, however, used either the peak proton flux of a specific energy range or the integrated X-ray flux. This can be explained by Big Flare syndrome mechanism: The statistically expected magnitude of measured flare energy manifestation is proportional to the energy released in a solar flare \citep{kahler1982role}.

\par The reason for the double power-law characteristics of the SEP spectrum has not been clearly understood to this day. The understanding of transition energy or break energy is crucial for this topic. Based on correlation between break energy and charge-to-mass ratio, several numerical simulation studies suggest that the double power-law spectrum of SEP events can be directly generated by the acceleration at the shock \cite[e.g.,][]{mason2012interplanetary,Desai_2016spectral_I,yu2022double}. Alternatively, \cite{mason2012interplanetary} set the initial energy spectrum as a double power-law and found that as particles propagate, the transition energy of the spectrum decreases. Other numerical studies even suggest that the double power-law spectrum can be naturally built from an initial single power-law spectrum through particle propagation processes\cite[e.g.,][]{Li_2015,zhao2016double}.

\par By linking the peak flux spectrum which is closely related to the accelerated particle spectrum with the integral spectrum which contains more propagation effect, we can gain a better understanding of the acceleration and propagation behavior of particles. 
Based on the statistical analysis of 103 fitted SEP peak spectra and 164 fluence spectra, our study investigates the correlation between the transition energy of the double power-law spectrum and the particle diffusion coefficient (see Sect. \ref{sec:Connections of two spectra}). We note that the transition energy of the fitted Band function has a slight dependence on the diffusion coefficient, indicating that SEP propagation process can contribute to the formation and/or evolution of the double power law.  

\par Moreover, as shown in Sect. \ref{sec:Distribution of parameters}, the three parameters in the Band function, logC, $\gamma_1$, and log$E_0$ approximately follow a normal distribution. We do not consider this to be a coincidence, but rather an indication that the physical processes determining these three parameters are different from the other parameters such as $\alpha$, $\kappa_0$, and $\gamma$. Generally speaking, the time-integral energy spectrum can be considered as the superposition of a series of instantaneous energy spectra. During a single SEP event, the solar wind parameters in space can be approximated as constant, allowing the parameters of this series of instantaneous energy spectra to be considered to follow the same distribution. According to the central limit theorem, the superposition of these parameters with the same distribution will follow a normal distribution. 

\par Overall, we outline the mechanism for establishing the double power-law spectrum of particle propagation process as follows: During the descending phase of the SEP event, particles collected by the detector may have traveled a much longer distance (including the multiple-crossing process) than the particles which arrived during the early phase due to the spatial propagation effect \citep{wang2023crucial}. Thus, during the descending phase transport effects including adiabatic cooling could significantly impact the spectra of SEPs as shown by simulations \citep{qin2006effect,mason2012interplanetary}. Since low-energy particles experience greater relative energy loss compared to high-energy particles, there would be more flattening of the energy spectrum in the low-energy region, resulting in a double power law. 
\par Further analysis and attribution of SEP properties due to either flare- and CME-acceleration mechanisms would help us better understand the acceleration nature and constrain of SEPs. Utilizing data collected at various solar distances rather than 1 AU would contribute to a better understanding of the transport effects and the spatial evolution. In order to fully understand SEP events, we require more comprehensive observation instruments that encompass a wider range of space, time, and energy.

\section*{Acknowledgments}
The authors acknowledge the support by the Strategic Priority Program of the Chinese Academy of Sciences (Grant No.XDB41000000 and ZDBS-SSW-TLC00103), the National Natural Science Foundation of China (Grant Nos. 42188101, 42074222, 42130204). The authors also express their gratitude to the groups developing SEPEM project for the continuous and high-quality dataset and thank Dr. Piers Jiggens for his help in accessing the newest version of the SEPEM dataset. 

\newpage



\begin{longrotatetable}
\setlength\tabcolsep{1.8pt}{ 
\begin{deluxetable*}{cccccccccccccccc}
\tablecaption{Peak flux spectral parameters fitted for 103 events.\label{tab:peak_flux_parameters}}
\tabletypesize{\footnotesize}
\tablewidth{1.0\textwidth}
\tablehead{
\colhead{No.} &   \colhead{No.\tablenotemark{$^{1}$}}&   \colhead{Start Time}     &   \colhead{End Time}      &   \colhead{$j_0$ \tablenotemark{$^{2}$}} &   \colhead{$j_0$\_err\tablenotemark{$^{2}$}} &\colhead{$\alpha$} &     \colhead{$\alpha$\_err} &   \colhead{$\kappa_0$\_450\tablenotemark{$^{3}$} }&   \colhead{$\kappa_0$\tablenotemark{$^{4}$}} &   \colhead{$\kappa_0$\_err\tablenotemark{$^{4}$}}&   \colhead{$\gamma$ }&   \colhead{$\gamma$\_err }&   \colhead{Vsw\tablenotemark{$^{5}$} }& \colhead{Vsw\_err\tablenotemark{$^{5}$}} & \colhead{$\chi^{2}$}  \\
}
\startdata
        1 & 2 & 1974-07-03 01:00 & 1974-07-10 13:30 & 4.02E+07 & 2.91E+07 & 1.49 & 0.33 & 2.41E+19 & 2.41E+19 & 1.72E+19 & 5 & 0.15 & 571.96 & 122.74 & 24.6 \\
        2 & 9 & 1977-09-18 16:30 & 1977-09-23 07:30 & 1.32E+06 & 1.27E+06 & 0.93 & 0.2 & 5.26E+19 & 5.26E+19 & 2.22E+19 & 3.43 & 0.16 & 407.1 & 40.59 & 3.81 \\
        3 & 11 & 1977-11-22 11:00 & 1977-11-28 13:30 & 2.72E+06 & 2.46E+06 & 1.21 & 0.15 & 2.87E+19 & 8.24E+18 & 3.28E+18 & 3.31 & 0.16 & 351.04 & 65.76 & 6.15 \\
        4 & 14 & 1978-02-13 08:00 & 1978-02-22 21:30 & 3.40E+06 & 5.38E+05 & 2.81 & 0.87 & 3.24E+10 & 3.10E+18 & 5.80E+18 & 4.03 & 0.04 & 412.58 & 77.81 & 20.6 \\
        5 & 15 & 1978-04-08 04:00 & 1978-04-16 03:00 & 1.10E+08 & 3.49E+08 & 0.72 & 0.14 & 3.45E+19 & 3.45E+19 & 9.17E+18 & 4.3 & 0.45 & 491.45 & 76.29 & 16.31 \\
        6 & 16 & 1978-04-17 03:30 & 1978-05-05 02:30 & 3.69E+06 & 5.96E+05 & 3.36 & 0.87 & 6.84E+17 & 6.84E+17 & 1.27E+18 & 4.05 & 0.04 & 514.36 & 109.31 & 33.4 \\
        7 & 21 & 1978-09-23 11:30 & 1978-10-16 01:30 & 6.73E+05 & 7.60E+04 & 3.58 & 1.2 & 4.96E+17 & 4.96E+17 & 1.27E+18 & 3.2 & 0.03 & 544.32 & 150.63 & 3.46 \\
        8 & 24 & 1979-02-17 16:30 & 1979-02-23 08:00 & 1.07E+06 & 1.33E+06 & 0.95 & 0.14 & 1.08E+19 & 2.43E+19 & 7.47E+18 & 3.7 & 0.2 & 509.62 & 66.95 & 11.23 \\
        9 & 30 & 1979-09-08 12:00 & 1979-10-03 04:30 & 1.99E+05 & 8.68E+04 & 1.34 & 0.23 & 2.26E+19 & 2.26E+19 & 1.17E+19 & 3.52 & 0.08 & 322.92 & 16.64 & 6.87 \\
        10 & 34 & 1980-02-05 18:30 & 1980-02-09 18:30 & 5.83E+03 & 1.10E+03 & 2.23 & 0.51 & 5.36E+18 & 5.36E+18 & 6.11E+18 & 3.13 & 0.04 & 426.83 & 51.93 & 6.41 \\
        11 & 35 & 1980-03-29 19:30 & 1980-04-08 20:00 & 1.35E+04 & 1.21E+04 & 1.04 & 0.19 & 2.03E+18 & 2.75E+19 & 1.22E+19 & 3.08 & 0.15 & 400.85 & 66.06 & 5.77 \\
        12 & 42 & 1981-04-24 04:00 & 1981-04-26 11:00 & 1.75E+05 & 7.31E+04 & 1.48 & 0.24 & 1.20E+19 & 1.20E+19 & 7.04E+18 & 3.18 & 0.08 & 477.98 & 66.89 & 18.84 \\
        13 & 43 & 1981-07-20 15:00 & 1981-07-28 07:00 & 7.40E+05 & 1.07E+06 & 0.78 & 0.2 & 8.76E+18 & 7.58E+19 & 3.42E+19 & 3.62 & 0.22 & 494.18 & 124.13 & 11.15 \\
        14 & 48 & 1981-10-08 03:30 & 1981-10-26 22:30 & 1.14E+06 & 8.33E+05 & 0.91 & 0.61 & 2.51E+20 & 2.51E+20 & 3.10E+20 & 3.38 & 0.12 & 439.16 & 71.41 & 5.88 \\
        15 & 51 & 1981-12-05 17:00 & 1981-12-14 08:00 & 2.23E+05 & 3.54E+04 & 6.91 & 10.46 & 4.85E+19 & 2.66E+15 & 5.39E+16 & 4.18 & 0.05 & 353.87 & 64.71 & 0.11 \\
        16 & 66 & 1984-03-07 01:00 & 1984-03-23 07:00 & 1.67E+04 & 9.78E+03 & 0.79 & 0.17 & 2.10E+19 & 9.27E+19 & 3.64E+19 & 2.73 & 0.08 & 525.95 & 44.12 & 1.93 \\
        17 & 67 & 1984-04-25 09:30 & 1984-05-15 19:30 & 3.78E+06 & 5.00E+05 & 2.97 & 0.51 & 5.13E+19 & 9.53E+17 & 1.08E+18 & 3.86 & 0.03 & 534 & 11.86 & 43.56 \\
        18 & 71 & 1985-07-09 02:30 & 1985-07-12 13:30 & 5.33E+04 & 1.25E+04 & 1.9 & 0.27 & 8.03E+19 & 4.90E+18 & 3.32E+18 & 3.21 & 0.05 & 471.66 & 57.1 & 9.26 \\
        19 & 73 & 1986-02-14 10:55 & 1986-02-19 07:40 & 1.06E+05 & 1.27E+04 & 3.26 & 0.63 & 1.38E+19 & 4.88E+17 & 6.90E+17 & 3.37 & 0.03 & 479.32 & 45.99 & 5.88 \\
        20 & 75 & 1986-05-04 12:20 & 1986-05-05 13:15 & 2.47E+05 & 7.16E+05 & 0.75 & 0.22 & 2.27E+19 & 5.45E+19 & 1.85E+19 & 3.75 & 0.45 & 416.78 & 21.79 & 1.35 \\
        21 & 76 & 1987-11-07 22:45 & 1987-11-10 13:15 & 4.65E+07 & 7.61E+07 & 1.3 & 0.25 & 1.75E+19 & 1.59E+19 & 8.25E+18 & 5.24 & 0.33 & 530.58 & 107.19 & 0.17 \\
        22 & 80 & 1988-11-08 15:45 & 1988-11-10 11:40 & 1.47E+02 & 1.28E+01 & 4.28 & 2.82 & 1.86E+17 & 1.86E+17 & 1.09E+18 & 2.25 & 0.02 & 416.87 & 55 & 7.21 \\
        23 & 83 & 1989-03-08 03:30 & 1989-03-14 19:50 & 3.32E+10 & 2.32E+10 & 1.57 & 0.19 & 1.08E+19 & 1.08E+19 & 5.34E+18 & 6.42 & 0.15 & 543.96 & 152.17 & 78.63 \\
        24 & 84 & 1989-03-23 20:15 & 1989-03-24 21:30 & 1.37E+04 & 5.54E+03 & 1.2 & 0.78 & 1.62E+20 & 1.62E+20 & 2.71E+20 & 3.23 & 0.07 & 533.33 & 83.85 & 20.8 \\
        25 & 85 & 1989-04-10 21:15 & 1989-04-18 01:45 & 2.50E+09 & 6.33E+09 & 1.08 & 0.23 & 2.72E+19 & 2.72E+19 & 1.07E+19 & 6.1 & 0.48 & 402.14 & 30.24 & 1.66 \\
        26 & 89 & 1989-07-25 09:05 & 1989-07-26 17:45 & 5.90E+02 & 1.26E+02 & 1.37 & 0.24 & 1.66E+19 & 1.66E+19 & 9.97E+18 & 2.07 & 0.04 & 432.62 & 36.96 & 5.74 \\
        27 & 90 & 1989-08-12 15:45 & 1989-08-14 15:45 & 3.02E+07 & 4.87E+06 & 2.1 & 0.16 & 1.48E+18 & 1.48E+18 & 6.96E+17 & 3.87 & 0.03 & 530.51 & 92.08 & 42.99 \\
        28 & 92 & 1989-09-29 11:55 & 1989-10-10 05:20 & 1.30E+05 & 2.30E+04 & 1.55 & 0.26 & 1.32E+19 & 1.32E+19 & 8.74E+18 & 2.39 & 0.03 & 394.27 & 25.07 & 8.29 \\
        29 & 93 & 1989-10-19 13:10 & 1989-11-09 16:50 & 1.15E+07 & 1.12E+06 & 2.68 & 0.6 & 1.55E+20 & 2.49E+18 & 3.44E+18 & 3.15 & 0.02 & 609.32 & 150.42 & 12.75 \\
        30 & 95 & 1989-11-27 06:25 & 1989-12-05 09:05 & 5.52E+08 & 1.37E+08 & 2.7 & 0.24 & 4.64E+17 & 4.64E+17 & 2.74E+17 & 5.2 & 0.05 & 554.5 & 85.57 & 90.49 \\
        31 & 96 & 1990-03-19 06:30 & 1990-03-22 01:40 & 1.31E+10 & 8.51E+09 & 1.78 & 0.14 & 2.67E+18 & 2.67E+18 & 1.08E+18 & 6.28 & 0.13 & 447.1 & 96.56 & 60.07 \\
        32 & 102 & 1990-05-17 21:30 & 1990-05-24 19:00 & 2.09E+04 & 2.36E+03 & 1.93 & 2.05 & 7.12E+19 & 7.13E+19 & 3.26E+20 & 2.57 & 0.02 & 418.11 & 62.97 & 4.31 \\
        33 & 106 & 1990-07-31 15:25 & 1990-08-06 12:05 & 2.52E+12 & 1.84E+13 & 0.83 & 0.2 & 2.47E+19 & 3.03E+19 & 7.69E+18 & 7.28 & 1.21 & 441 & 60.6 & 0.88 \\
        34 & 108 & 1991-01-27 14:45 & 1991-02-02 19:30 & 1.14E+07 & 2.79E+06 & 3.45 & 0.48 & 8.68E+18 & 2.19E+17 & 2.32E+17 & 5.02 & 0.06 & 376.37 & 33.78 & 3.28 \\
        35 & 110 & 1991-02-25 10:40 & 1991-02-27 01:55 & 8.17E+03 & 1.57E+03 & 3.92 & 0.89 & 5.58E+18 & 1.08E+17 & 2.07E+17 & 3.46 & 0.05 & 422.79 & 48.27 & 47.11 \\
        36 & 111 & 1991-03-12 18:50 & 1991-03-13 22:25 & 2.50E+05 & 5.50E+05 & 1.15 & 0.32 & 2.34E+19 & 2.34E+19 & 1.37E+19 & 4.19 & 0.43 & 363.9 & 27.61 & 0.34 \\
        37 & 113 & 1991-03-23 06:40 & 1991-03-31 14:30 & 2.87E+08 & 4.12E+07 & 2.49 & 0.21 & 1.13E+19 & 8.15E+17 & 4.39E+17 & 4.12 & 0.03 & 478.67 & 13.01 & 61.51 \\
        38 & 116 & 1991-05-10 15:05 & 1991-05-15 10:50 & 5.20E+04 & 4.11E+03 & 3.9 & 1.24 & 7.11E+19 & 2.47E+17 & 6.51E+17 & 2.97 & 0.02 & 364.81 & 13.16 & 10.25 \\
        39 & 118 & 1991-06-09 19:10 & 1991-06-15 01:30 & 3.57E+06 & 6.14E+05 & 2.01 & 0.19 & 1.91E+18 & 1.91E+18 & 1.02E+18 & 3.42 & 0.03 & 541.21 & 73.29 & 18.18 \\
        40 & 120 & 1991-08-25 21:10 & 1991-08-30 22:30 & 1.66E+08 & 2.48E+08 & 1.26 & 0.27 & 5.86E+19 & 2.48E+19 & 1.36E+19 & 5.51 & 0.31 & 435.36 & 79.36 & 2.27 \\
        41 & 127 & 1992-02-06 22:45 & 1992-02-10 00:30 & 1.71E+07 & 1.20E+07 & 2.63 & 0.73 & 2.32E+19 & 2.45E+18 & 3.60E+18 & 5.57 & 0.19 & 461.75 & 58.05 & 0.61 \\
        42 & 130 & 1992-05-09 06:15 & 1992-05-13 20:15 & 6.84E+09 & 2.09E+09 & 2.29 & 0.21 & 1.91E+18 & 1.91E+18 & 1.13E+18 & 6.04 & 0.07 & 429.13 & 150.13 & 54.45 \\
        43 & 131 & 1992-06-25 20:30 & 1992-07-01 23:25 & 1.91E+04 & 1.58E+03 & 3.79 & 0.79 & 7.56E+18 & 1.10E+17 & 1.98E+17 & 2.61 & 0.02 & 536.66 & 41.41 & 10.18 \\
        44 & 133 & 1992-10-30 18:45 & 1992-11-01 21:20 & 5.29E+06 & 7.49E+05 & 4.19 & 0.49 & 6.46E+15 & 6.46E+15 & 8.09E+15 & 3.75 & 0.03 & 404.82 & 28.12 & 50.09 \\
        45 & 134 & 1993-03-04 13:20 & 1993-03-05 22:30 & 1.20E+03 & 2.31E+02 & 1.87 & 0.89 & 2.55E+19 & 2.55E+19 & 5.05E+19 & 2.64 & 0.04 & 416.61 & 77.49 & 7.2 \\
        46 & 136 & 1993-03-12 18:50 & 1993-03-14 15:05 & 2.98E+03 & 5.88E+02 & 1.69 & 0.89 & 4.50E+19 & 4.50E+19 & 8.95E+19 & 2.63 & 0.04 & 520.48 & 40.97 & 3.5 \\
        47 & 140 & 1997-11-04 06:50 & 1997-11-10 19:40 & 9.73E+08 & 9.04E+09 & 0.33 & 0.12 & 1.01E+20 & 1.19E+20 & 1.94E+19 & 3.59 & 0.82 & 371.21 & 38.3 & 3.84 \\
        48 & 141 & 1998-04-20 12:55 & 1998-04-26 15:05 & 4.52E+08 & 1.68E+08 & 2.76 & 0.19 & 1.19E+18 & 4.88E+16 & 2.76E+16 & 4.79 & 0.08 & 406.23 & 52.66 & 24.61 \\
        49 & 143 & 1998-05-02 13:55 & 1998-05-03 04:10 & 9.54E+03 & 1.70E+03 & 1.51 & 0.23 & 1.34E+19 & 1.34E+19 & 8.12E+18 & 2.47 & 0.03 & 547.41 & 85.11 & 3.7 \\
        50 & 144 & 1998-05-06 08:25 & 1998-05-08 00:20 & 8.13E+04 & 1.10E+04 & 3.7 & 0.47 & 1.54E+16 & 1.54E+16 & 1.95E+16 & 3.07 & 0.03 & 497.11 & 48.28 & 5.73 \\
        51 & 145 & 1998-05-09 06:50 & 1998-05-11 04:10 & 1.73E+05 & 2.25E+06 & 0.52 & 0.69 & 1.62E+20 & 1.62E+20 & 6.19E+19 & 3.7 & 1.77 & 426.97 & 74.82 & 2.12 \\
        52 & 149 & 1998-09-30 14:25 & 1998-10-04 04:20 & 7.04E+06 & 1.83E+06 & 2.21 & 0.23 & 1.57E+18 & 1.57E+18 & 9.53E+17 & 4.06 & 0.05 & 474.22 & 88.81 & 23.68 \\
        53 & 152 & 1998-11-14 06:30 & 1998-11-17 15:55 & 5.28E+05 & 2.14E+05 & 1.52 & 0.17 & 1.76E+19 & 6.62E+18 & 2.93E+18 & 3.41 & 0.08 & 432.73 & 51.39 & 12.13 \\
        54 & 156 & 1999-05-27 12:15 & 1999-05-28 14:40 & 8.15E+04 & 2.70E+05 & 1.02 & 0.36 & 3.29E+19 & 2.62E+19 & 1.51E+19 & 3.72 & 0.61 & 388.59 & 35.9 & 0.2 \\
        55 & 158 & 2000-02-18 08:45 & 2000-02-19 11:55 & 2.33E+03 & 8.73E+02 & 1.45 & 0.43 & 2.94E+19 & 2.94E+19 & 2.95E+19 & 2.88 & 0.07 & 436.73 & 136.37 & 20.87 \\
        56 & 161 & 2000-06-10 04:00 & 2000-06-12 21:10 & 3.68E+04 & 3.66E+04 & 0.86 & 0.22 & 6.65E+19 & 6.65E+19 & 3.15E+19 & 3.09 & 0.16 & 540.76 & 84.83 & 11.98 \\
        57 & 163 & 2000-07-13 00:30 & 2000-07-23 19:20 & 1.57E+07 & 2.03E+06 & 2.33 & 0.2 & 6.29E+17 & 6.29E+17 & 3.81E+17 & 3.22 & 0.02 & 661.16 & 135.08 & 11.81 \\
        58 & 166 & 2000-09-12 14:50 & 2000-09-18 01:25 & 1.07E+06 & 1.38E+05 & 5.13 & 1.45 & 1.02E+18 & 1.82E+16 & 5.45E+16 & 4.19 & 0.03 & 478.27 & 154 & 2.53 \\
        59 & 168 & 2000-10-25 15:15 & 2000-10-27 19:10 & 7.27E+03 & 9.58E+02 & 6.72 & 58.72 & 2.19E+16 & 2.19E+16 & 2.48E+18 & 3.42 & 0.04 & 377.28 & 32.77 & 0.61 \\
        60 & 170 & 2000-11-08 23:45 & 2000-11-15 17:00 & 2.84E+11 & 1.43E+11 & 1.27 & 0.05 & 4.05E+18 & 3.14E+18 & 9.34E+17 & 5.21 & 0.08 & 605.41 & 151.47 & 5.03 \\
        61 & 171 & 2000-11-24 07:00 & 2000-11-25 20:10 & 1.63E+04 & 1.88E+03 & 3.18 & 1.2 & 1.34E+18 & 1.34E+18 & 3.44E+18 & 3.01 & 0.03 & 497.56 & 71.98 & 10.93 \\
        62 & 176 & 2001-04-02 11:20 & 2001-04-07 08:20 & 1.42E+07 & 3.18E+06 & 3.38 & 0.3 & 1.57E+16 & 1.57E+16 & 1.38E+16 & 4.1 & 0.05 & 548.41 & 96.53 & 19.49 \\
        63 & 178 & 2001-05-07 15:00 & 2001-05-09 19:10 & 2.73E+04 & 4.38E+03 & 5.25 & 2.29 & 1.96E+16 & 1.96E+16 & 9.11E+16 & 3.67 & 0.04 & 474.33 & 80.86 & 4.02 \\
        64 & 179 & 2001-05-20 08:45 & 2001-05-21 14:25 & 1.46E+02 & 2.16E+01 & 2.84 & 0.61 & 5.14E+17 & 5.14E+17 & 7.61E+17 & 2.13 & 0.03 & 417.18 & 87.24 & 10.12 \\
        65 & 182 & 2001-08-16 00:55 & 2001-08-26 04:40 & 1.21E+07 & 2.45E+06 & 2.4 & 0.14 & 3.28E+17 & 7.03E+16 & 3.44E+16 & 3.75 & 0.04 & 459.33 & 77.22 & 8.39 \\
        66 & 183 & 2001-09-24 12:00 & 2001-10-12 03:05 & 1.13E+08 & 1.96E+07 & 2.23 & 0.2 & 1.61E+18 & 1.61E+18 & 8.27E+17 & 4.2 & 0.03 & 508.35 & 76.6 & 20.16 \\
        67 & 185 & 2001-11-05 16:55 & 2001-11-12 20:05 & 1.11E+10 & 3.20E+09 & 1.43 & 0.07 & 4.01E+18 & 4.01E+18 & 1.16E+18 & 4.69 & 0.05 & 482.65 & 98.21 & 8.26 \\
        68 & 186 & 2001-11-17 19:55 & 2001-11-30 13:00 & 1.01E+11 & 6.11E+10 & 1.69 & 0.13 & 3.46E+18 & 3.46E+18 & 1.52E+18 & 5.97 & 0.12 & 460.88 & 141.35 & 71.59 \\
        69 & 188 & 2001-12-26 05:55 & 2002-01-09 07:00 & 2.29E+06 & 9.54E+05 & 1.15 & 0.09 & 4.08E+18 & 1.04E+19 & 3.11E+18 & 3.23 & 0.07 & 416.91 & 65.95 & 7.92 \\
        70 & 189 & 2002-01-10 09:55 & 2002-01-18 18:35 & 4.76E+05 & 1.44E+05 & 2.7 & 0.56 & 8.15E+18 & 1.72E+18 & 2.10E+18 & 4.25 & 0.08 & 486.11 & 86.35 & 1.1 \\
        71 & 191 & 2002-04-17 11:30 & 2002-04-28 13:35 & 5.53E+07 & 1.48E+07 & 1.75 & 0.12 & 8.16E+18 & 2.53E+18 & 8.57E+17 & 4.12 & 0.05 & 492.63 & 56.64 & 29.08 \\
        72 & 192 & 2002-05-22 07:50 & 2002-05-25 00:15 & 3.10E+07 & 5.56E+06 & 3.98 & 1.5 & 2.15E+18 & 3.86E+17 & 1.18E+18 & 5.1 & 0.05 & 555.97 & 135.26 & 1.35 \\
        73 & 193 & 2002-07-07 14:00 & 2002-07-09 13:55 & 1.26E+05 & 1.41E+05 & 1.61 & 0.31 & 5.35E+18 & 6.03E+18 & 3.99E+18 & 3.86 & 0.24 & 431.76 & 44.02 & 1.08 \\
        74 & 200 & 2002-11-09 17:10 & 2002-11-11 21:15 & 3.29E+07 & 1.84E+07 & 2.06 & 0.29 & 1.17E+19 & 3.27E+18 & 2.13E+18 & 5.05 & 0.13 & 487.19 & 81.09 & 1.37 \\
        75 & 205 & 2003-10-28 03:30 & 2003-10-29 16:50 & 2.04E+08 & 3.53E+07 & 1.92 & 0.16 & 2.99E+18 & 2.99E+18 & 1.55E+18 & 3.96 & 0.03 & 663.49 & 204.97 & 21.88 \\
        76 & 207 & 2003-12-02 12:50 & 2003-12-06 06:05 & 1.46E+08 & 1.18E+08 & 1.7 & 0.24 & 7.66E+18 & 7.66E+18 & 4.24E+18 & 5.79 & 0.18 & 446.6 & 89.46 & 2.07 \\
        77 & 209 & 2004-07-23 15:05 & 2004-07-28 18:10 & 1.68E+10 & 8.63E+09 & 2.32 & 0.22 & 1.36E+18 & 1.36E+18 & 7.55E+17 & 6.48 & 0.12 & 623.09 & 141.75 & 4.77 \\
        78 & 211 & 2004-09-13 19:55 & 2004-09-17 18:15 & 2.28E+07 & 9.11E+06 & 2.36 & 0.47 & 3.97E+18 & 3.97E+18 & 4.03E+18 & 5.08 & 0.1 & 484.2 & 71.98 & 7.68 \\
        79 & 213 & 2004-11-01 06:10 & 2004-11-02 20:15 & 7.68E+04 & 2.56E+04 & 1.94 & 0.22 & 1.91E+18 & 1.91E+18 & 1.16E+18 & 3.3 & 0.07 & 399.63 & 86.18 & 14.57 \\
        80 & 214 & 2004-11-07 02:50 & 2004-11-09 02:45 & 5.25E+06 & 7.98E+05 & 3.24 & 0.46 & 4.03E+17 & 4.03E+17 & 4.25E+17 & 4.37 & 0.03 & 597.56 & 106.37 & 29.61 \\
        81 & 216 & 2005-01-20 01:00 & 2005-01-21 00:00 & 2.40E+04 & 1.57E+04 & 0.77 & 0.24 & 8.12E+19 & 8.12E+19 & 4.73E+19 & 1.74 & 0.09 & 694.59 & 152.67 & 9.11 \\
        82 & 218 & 2005-05-13 21:00 & 2005-05-17 12:30 & 8.58E+10 & 3.64E+10 & 3.04 & 0.3 & 3.69E+17 & 3.69E+17 & 2.66E+17 & 7.16 & 0.11 & 559.26 & 137.71 & 17.48 \\
        83 & 219 & 2005-06-16 20:50 & 2005-06-18 05:55 & 7.78E+03 & 2.65E+03 & 1.61 & 0.23 & 5.53E+18 & 5.53E+18 & 3.35E+18 & 2.68 & 0.06 & 430.04 & 92.27 & 9.3 \\
        84 & 220 & 2005-07-13 18:10 & 2005-07-20 01:55 & 1.65E+12 & 1.18E+13 & 1.02 & 0.19 & 1.05E+19 & 1.05E+19 & 2.82E+18 & 7.01 & 1.25 & 443.63 & 53.29 & 1.56 \\
        85 & 221 & 2005-07-26 23:20 & 2005-08-04 21:40 & 3.09E+06 & 3.65E+07 & 0.49 & 0.83 & 3.17E+20 & 3.17E+20 & 1.61E+20 & 4.39 & 1.58 & 489.66 & 78.23 & 2.94 \\
        86 & 224 & 2006-12-05 17:35 & 2006-12-11 08:00 & 2.08E+06 & 3.13E+05 & 2.13 & 0.31 & 4.33E+18 & 4.33E+18 & 3.23E+18 & 3.45 & 0.03 & 585.02 & 90.67 & 12.46 \\
        87 & 227 & 2011-03-07 23:15 & 2011-03-12 07:20 & 1.00E+09 & 6.67E+09 & 0.81 & 0.27 & 3.94E+19 & 3.94E+19 & 1.56E+19 & 5.64 & 1.11 & 438.2 & 100.95 & 4.95 \\
        88 & 229 & 2011-06-05 20:05 & 2011-06-12 05:30 & 4.62E+04 & 4.18E+04 & 0.89 & 0.21 & 5.40E+19 & 5.40E+19 & 2.38E+19 & 2.96 & 0.14 & 438 & 39.49 & 9.24 \\
        89 & 231 & 2011-08-09 08:00 & 2011-08-10 16:20 & 4.42E+04 & 5.37E+04 & 0.85 & 0.17 & 3.71E+19 & 3.71E+19 & 1.42E+19 & 3 & 0.19 & 449.37 & 77.31 & 13.15 \\
        90 & 232 & 2011-09-06 23:45 & 2011-09-08 13:35 & 5.64E+02 & 1.31E+02 & 2.01 & 0.36 & 3.55E+18 & 3.55E+18 & 3.21E+18 & 2.51 & 0.05 & 491.28 & 106.18 & 10.22 \\
        91 & 236 & 2011-11-26 09:00 & 2011-11-30 00:20 & 2.42E+06 & 9.56E+05 & 2.33 & 0.5 & 4.18E+18 & 4.18E+18 & 4.45E+18 & 4.7 & 0.1 & 415.88 & 39.84 & 7.18 \\
        92 & 237 & 2012-01-26 11:25 & 2012-02-03 03:55 & 2.30E+05 & 2.05E+04 & 4.57 & 0.75 & 7.87E+15 & 7.87E+15 & 1.41E+16 & 3.1 & 0.02 & 437.29 & 93.44 & 13.06 \\
        93 & 239 & 2012-03-05 00:30 & 2012-03-17 02:25 & 6.75E+06 & 8.57E+05 & 1.8 & 0.34 & 1.70E+19 & 1.70E+19 & 1.43E+19 & 3.48 & 0.02 & 491.29 & 115.42 & 2.13 \\
        94 & 240 & 2012-05-17 02:05 & 2012-05-20 15:15 & 1.64E+05 & 2.17E+05 & 0.56 & 0.18 & 1.67E+20 & 1.67E+20 & 6.56E+19 & 2.8 & 0.16 & 417.18 & 84.53 & 15.84 \\
        95 & 244 & 2012-07-12 18:05 & 2012-07-13 13:00 & 1.40E+06 & 3.75E+05 & 4.32 & 0.58 & 1.10E+16 & 1.10E+16 & 1.53E+16 & 4.48 & 0.07 & 443.89 & 84.35 & 12.58 \\
        96 & 247 & 2012-09-28 01:20 & 2012-10-01 03:35 & 1.90E+05 & 1.02E+05 & 2.6 & 0.35 & 2.15E+17 & 3.83E+17 & 3.23E+17 & 3.98 & 0.13 & 324.32 & 30.3 & 2.9 \\
        97 & 250 & 2013-04-11 09:25 & 2013-04-14 17:00 & 4.23E+04 & 7.50E+03 & 2.56 & 0.36 & 8.32E+17 & 8.32E+17 & 7.44E+17 & 3.13 & 0.04 & 406.66 & 55.84 & 12.57 \\
        98 & 251 & 2013-05-14 07:20 & 2013-05-21 15:20 & 2.98E+05 & 7.13E+04 & 3.12 & 0.8 & 1.07E+18 & 1.07E+18 & 1.80E+18 & 4.23 & 0.06 & 388.82 & 24.88 & 3.12 \\
        99 & 252 & 2013-05-22 13:45 & 2013-05-25 23:35 & 8.41E+06 & 1.28E+06 & 4.72 & 0.62 & 3.17E+15 & 3.17E+15 & 4.83E+15 & 4.12 & 0.03 & 574.21 & 110.62 & 33.82 \\
        100 & 260 & 2015-06-18 05:05 & 2015-07-02 17:05 & 9.13E+08 & 2.41E+08 & 3.33 & 0.42 & 3.41E+17 & 3.42E+17 & 3.31E+17 & 5.99 & 0.07 & 447.6 & 139.7 & 10.58 \\
        101 & 261 & 2016-01-02 01:05 & 2016-01-03 12:55 & 3.31E+04 & 6.53E+03 & 3.85 & 1.85 & 4.83E+17 & 4.83E+17 & 1.82E+18 & 3.75 & 0.05 & 489.53 & 61.4 & 10.46 \\
        102 & 262 & 2017-07-14 04:50 & 2017-07-15 00:00 & 5.16E+04 & 1.24E+04 & 3.04 & 1.1 & 1.78E+18 & 1.78E+18 & 4.10E+18 & 3.84 & 0.06 & 437.96 & 87.08 & 2.17 \\
        103 & 263 & 2017-09-04 22:30 & 2017-09-15 12:40 & 1.07E+15 & 3.09E+16 & 0.24 & 0.12 & 1.32E+20 & 1.32E+20 & 5.85E+19 & 5.03 & 2.04 & 560.26 & 94.12 & 7.84\\
\enddata
\end{deluxetable*}}
\vspace{-26pt}
\tablenotetext{1}{Number in SEPEM reference event list.}
\tablenotetext{2}{The normalization coefficient for Equation \ref{eqn:peak_spectrum} and its uncertainty are expressed in units of $[\#/cm^2/s/sr/MeV]$.}
\tablenotetext{3}{The normalized diffusion coefficient for Equation \ref{eqn:peak_spectrum} which is calculated using a constant solar wind speed of 450 $[km/s]$. Expressed in units of $[cm^2/s]$.}
\tablenotetext{4}{The normalized diffusion coefficient for Equation \ref{eqn:peak_spectrum} which is calculated using a real-time solar wind speed list in this table. Expressed in units of $[cm^2/s]$.}
\tablenotetext{5}{The real-time solar wind speed for Equation \ref{eqn:peak_spectrum} and its uncertainty are expressed in units of $[km/s]$.}
\end{longrotatetable}

\begin{longrotatetable}
\setlength\tabcolsep{2.3pt}{ 
\begin{deluxetable*}{ccccccccccccccccc}
\tablecaption{Integral fluence spectral parameters fitted for 164 events. \label{tab:Integral_fluence_parameters}}
\tabletypesize{\footnotesize}
\tablewidth{1.0\textwidth}
\tablehead{
    \colhead{No.} &   \colhead{No.\tablenotemark{$^{1}$} }&   \colhead{Start Time}     &   \colhead{End Time}      &  \colhead{C \tablenotemark{$^{2}$}} &  \colhead{$\gamma_1$}  &  \colhead{$\gamma_2 $}  &  \colhead{$E_0$ \tablenotemark{$^{3}$}} &  \colhead{$E_{break}$ \tablenotemark{$^{4}$}}  &  \colhead{C\_err\tablenotemark{$^{2}$}} &  \colhead{$\gamma_1$\_err}  &  \colhead{$\gamma_2$\_err} &  \colhead{$E_0$\_err \tablenotemark{$^{3}$}} &  \colhead{$E_{break}$\_err \tablenotemark{$^{4}$}}  & \colhead{GLE} & \colhead{flare}&\colhead{$\chi^{2}$}\\
}
\startdata
 1 & 2 & 1974-07-03 01:00 & 1974-07-10 13:30 & 4.62E+09 & -2.3 & -5.57 & 9.44 & 30.85 & 1.33E+09 & 0.18 & 0.08 & 1.05 & 7.51 &  &  & 31.24 \\
        2 & 3 & 1974-09-11 00:00 & 1974-10-03 18:00 & 1.98E+09 & -2.04 & -3.5 & 8.94 & 13.1 & 1.93E+09 & 0.83 & 0.02 & 7.18 & 33.43 &  &  & 23.20 \\
        3 & 4 & 1974-11-05 07:00 & 1974-11-08 19:00 & 2.13E+07 & -2.21 & -2.98 & 52.04 & 39.76 & 4.44E+06 & 0.12 & 0.03 & 16.83 & 66.78 &  &  & 1.03 \\
        4 & 5 & 1975-08-21 17:00 & 1975-08-24 05:00 & 3.34E+06 & -2.06 & -3.61 & 57.53 & 89.03 & 4.42E+05 & 0.06 & 0.07 & 6.53 & 31.15 &  &  & 5.09 \\
        5 & 6 & 1976-04-30 22:00 & 1976-05-05 03:30 & 5.78E+07 & -1.95 & -2.95 & 116.25 & 116.12 & 6.61E+06 & 0.05 & 0.09 & 16.8 & 67.22 & Yes & X2 & 2.29 \\
        6 & 7 & 1976-08-22 16:30 & 1976-08-24 23:00 & 4.50E+06 & -1.89 & -3.89 & 19.36 & 38.79 & 1.03E+06 & 0.13 & 0.08 & 2.71 & 13.71 &  & M3 & 18.03 \\
        7 & 9 & 1977-09-18 16:30 & 1977-09-23 07:30 & 4.78E+08 & -1.93 & -3.32 & 32.79 & 45.45 & 9.03E+07 & 0.1 & 0.04 & 5.4 & 23.08 & Yes & X2 & 1.62 \\
        8 & 11 & 1977-11-22 11:00 & 1977-11-28 13:30 & 6.13E+07 & -1.23 & -3.12 & 27.8 & 52.3 & 1.05E+07 & 0.09 & 0.04 & 3 & 11.64 & Yes & X1 & 1.83 \\
        9 & 12 & 1978-01-02 07:30 & 1978-01-06 16:00 & 7.58E+06 & -1.89 & -3.74 & 22.16 & 40.95 & 1.90E+06 & 0.14 & 0.04 & 3.39 & 16.15 &  & C7 & 22.45 \\
        10 & 14 & 1978-02-13 08:00 & 1978-02-22 21:30 & 1.76E+10 & -2.67 & -4.46 & 13.26 & 23.78 & 6.17E+09 & 0.24 & 0.04 & 3.23 & 18.83 &  & M7 & 43.40 \\
        11 & 15 & 1978-04-08 04:00 & 1978-04-16 03:00 & 6.49E+07 & -1.85 & -3.48 & 29.86 & 48.62 & 1.23E+07 & 0.1 & 0.04 & 4.08 & 18.09 &  & X2 & 3.13 \\
        12 & 16 & 1978-04-17 03:30 & 1978-05-05 02:30 & 1.14E+10 & -2.37 & -3.74 & 14.18 & 19.43 & 5.21E+09 & 0.33 & 0.03 & 5.69 & 28.06 & Yes & X5 & 16.67 \\
        13 & 21 & 1978-09-23 11:30 & 1978-10-16 01:30 & 4.32E+09 & -1.76 & -3.77 & 13.22 & 26.53 & 1.29E+09 & 0.19 & 0.03 & 2.37 & 11.31 &  & X1 & 16.71 \\
        14 & 24 & 1979-02-17 16:30 & 1979-02-23 08:00 & 1.40E+07 & -1.77 & -3.28 & 23.7 & 35.85 & 3.37E+06 & 0.14 & 0.03 & 4.31 & 18.07 &  & X2 & 4.37 \\
        15 & 27 & 1979-06-06 12:30 & 1979-06-14 10:00 & 3.84E+10 & -3.57 & -6.83 & 8.02 & 26.15 & 1.24E+10 & 0.21 & 0.14 & 1.02 & 9.35 &  & X2 & 21.14 \\
        16 & 29 & 1979-08-01 11:30 & 1979-09-01 08:00 & 2.30E+09 & -2.36 & -3.48 & 26.95 & 30.19 & 6.20E+08 & 0.17 & 0.03 & 7.8 & 35.85 & Yes & X6 & 5.66 \\
        17 & 30 & 1979-09-08 12:00 & 1979-10-03 04:30 & 1.83E+08 & -1.51 & -3.76 & 19.02 & 42.78 & 3.69E+07 & 0.11 & 0.04 & 2.04 & 9.56 &  & X2 & 4.40 \\
        18 & 31 & 1979-11-16 00:00 & 1979-11-23 12:00 & 9.97E+08 & -2.96 & -4.44 & 10.78 & 15.96 & 6.20E+08 & 0.49 & 0.03 & 5.48 & 32.55 &  & M1 & 36.32 \\
        19 & 34 & 1980-02-05 18:30 & 1980-02-09 18:30 & 6.79E+06 & -1.66 & -4.19 & 12.81 & 32.42 & 1.85E+06 & 0.16 & 0.04 & 1.65 & 8.67 &  & M3 & 32.65 \\
        20 & 35 & 1980-03-29 19:30 & 1980-04-08 20:00 & 3.01E+06 & -1.59 & -3.15 & 30.92 & 48.31 & 5.98E+05 & 0.1 & 0.04 & 4.58 & 18.33 &  & C5 & 4.26 \\
        21 & 36 & 1980-07-17 15:30 & 1980-07-26 18:30 & 5.92E+09 & -2.96 & -6.72 & 7.1 & 26.72 & 1.91E+09 & 0.21 & 0.16 & 0.8 & 7.22 &  & M3.4 & 25.99 \\
        22 & 39 & 1980-10-15 10:30 & 1980-10-24 16:30 & 7.76E+07 & -1.89 & -5.44 & 8.59 & 30.45 & 2.27E+07 & 0.18 & 0.16 & 0.91 & 6.71 &  & M2.1 & 10.58 \\
        23 & 40 & 1980-11-09 02:00 & 1980-12-05 08:30 & 9.31E+07 & -2.36 & -11.29 & 11.38 & 101.68 & 1.96E+07 & 0.12 & 2.15 & 0.74 & 32.67 &  & M7.2 & 106.87 \\
        24 & 41 & 1981-03-26 03:30 & 1981-04-07 20:30 & 1.43E+08 & -2.59 & -3.02 & 202.65 & 88.35 & 2.09E+07 & 0.07 & 0.07 & 83.3 & 345.35 &  & M3.5 & 5.90 \\
        25 & 42 & 1981-04-24 04:00 & 1981-04-26 11:00 & 1.19E+08 & -1.33 & -3.61 & 13.14 & 30 & 3.43E+07 & 0.18 & 0.04 & 2.03 & 9.16 & Yes & X5.9 & 34.01 \\
        26 & 43 & 1981-07-20 15:00 & 1981-07-28 07:00 & 3.79E+08 & -2.51 & -3.61 & 33.73 & 36.92 & 8.53E+07 & 0.13 & 0.04 & 8.17 & 38.96 &  & M5.4 & 6.51 \\
        27 & 48 & 1981-10-08 03:30 & 1981-10-26 22:30 & 3.35E+09 & -2.04 & -3.46 & 31.28 & 44.24 & 6.75E+08 & 0.11 & 0.04 & 5.27 & 23.54 & Yes & X3.6 & 4.71 \\
        28 & 51 & 1981-12-05 17:00 & 1981-12-14 08:00 & 1.80E+09 & -3.18 & -5.04 & 15.95 & 29.6 & 5.33E+08 & 0.19 & 0.14 & 3.28 & 22.39 &  & M5.2 & 3.34 \\
        29 & 53 & 1982-01-29 19:00 & 1982-02-12 01:00 & 1.81E+09 & -1.69 & -4.3 & 11.26 & 29.41 & 5.14E+08 & 0.18 & 0.04 & 1.51 & 8.13 &  & X1.1 & 29.49 \\
        30 & 54 & 1982-03-07 06:30 & 1982-03-10 08:30 & 4.04E+07 & -2.43 & -3.37 & 20.3 & 19.13 & 1.92E+07 & 0.34 & 0.03 & 12.07 & 54.61 &  &  & 12.12 \\
        31 & 56 & 1982-06-03 00:00 & 1982-07-02 06:00 & 9.82E+07 & -1.98 & -3.65 & 18.16 & 30.19 & 2.80E+07 & 0.18 & 0.04 & 3.81 & 17.88 &  & X12 & 33.11 \\
        32 & 57 & 1982-07-09 02:30 & 1982-07-27 05:00 & 6.21E+10 & -3.23 & -5.04 & 11.16 & 20.24 & 2.73E+10 & 0.32 & 0.04 & 3.37 & 22.43 &  & X9 & 69.02 \\
        33 & 60 & 1982-12-25 20:00 & 1983-01-10 00:30 & 3.52E+09 & -2.68 & -4.71 & 11.77 & 23.9 & 1.23E+09 & 0.24 & 0.04 & 2.54 & 15.52 &  & X2.2 & 53.49 \\
        34 & 61 & 1983-02-03 08:30 & 1983-02-10 04:00 & 1.97E+10 & -3.99 & -5.26 & 21.66 & 27.49 & 6.14E+09 & 0.2 & 0.05 & 6.62 & 46.89 &  & X4.1 & 73.02 \\
        35 & 63 & 1983-06-14 18:00 & 1983-06-29 15:00 & 5.41E+06 & -1.47 & -3.72 & 27.14 & 61.09 & 9.35E+05 & 0.09 & 0.04 & 2.46 & 11.35 &  & C2.6 & 5.31 \\
        36 & 66 & 1984-03-07 01:00 & 1984-03-23 07:00 & 9.56E+07 & -2.43 & -5.11 & 134.7 & 361.03 & 6.42E+06 & 0.02 & 0.16 & 6.57 & 54.9 &  & M2 & 29.19 \\
        37 & 67 & 1984-04-25 09:30 & 1984-05-15 19:30 & 2.58E+09 & -2.07 & -4.5 & 23.66 & 57.42 & 1.14E+08 & 0.03 & 2.07 & 1.6 & 50.09 &  & X13 & 5781.59 \\
        38 & 69 & 1985-01-22 02:30 & 1985-01-25 18:00 & 1.59E+07 & -2.37 & -3.98 & 56.09 & 90.79 & 2.23E+06 & 0.06 & 0.14 & 6.46 & 36.2 &  & X4.7 & 9.39 \\
        39 & 71 & 1985-07-09 02:30 & 1985-07-12 13:30 & 7.95E+06 & -1.49 & -3.72 & 19.05 & 42.48 & 1.63E+06 & 0.11 & 0.03 & 2.12 & 9.77 &  & M2.9 & 9.77 \\
        40 & 73 & 1986-02-14 10:55 & 1986-02-19 07:40 & 1.30E+08 & -1.31 & -3.79 & 9.35 & 23.24 & 4.62E+07 & 0.24 & 0.02 & 1.59 & 7.49 &  & M6.4 & 14.74 \\
        41 & 75 & 1986-05-04 12:20 & 1986-05-05 13:15 & 1.04E+06 & -1.23 & -4.56 & 18.12 & 60.39 & 1.72E+05 & 0.08 & 0.08 & 1.16 & 6.86 &  & M1.2 & 10.83 \\
        42 & 76 & 1987-11-07 22:45 & 1987-11-10 13:15 & 2.37E+08 & -2.16 & -6.33 & 9.56 & 39.93 & 5.37E+07 & 0.13 & 0.16 & 0.7 & 6.19 &  & M1.2 & 22.20 \\
        43 & 77 & 1987-12-30 02:15 & 1988-01-01 02:05 & 4.69E+07 & -2.22 & -3.83 & 9.49 & 15.29 & 3.36E+07 & 0.58 & 0.09 & 5.3 & 26.97 &  & C1.7 & 0.04 \\
        44 & 78 & 1988-01-02 23:00 & 1988-01-06 13:50 & 6.52E+08 & -2.43 & -5.27 & 17.82 & 50.66 & 1.20E+08 & 0.1 & 0.17 & 1.52 & 11.35 &  & X1.4 & 0.40 \\
        45 & 80 & 1988-11-08 15:45 & 1988-11-10 11:40 & 2.74E+06 & -1.58 & -3 & 31.06 & 44.15 & 5.26E+05 & 0.11 & 0.04 & 5.24 & 20.03 &  & M3 & 16.81 \\
        46 & 83 & 1989-03-08 03:30 & 1989-03-14 19:50 & 1.09E+10 & -2.73 & -4.92 & 14.76 & 32.2 & 2.89E+09 & 0.16 & 0.04 & 2.18 & 13.78 &  & X15 & 49.58 \\
        47 & 84 & 1989-03-23 20:15 & 1989-03-24 21:30 & 2.86E+07 & -2.14 & -4.69 & 21.4 & 54.58 & 4.87E+06 & 0.09 & 0.04 & 1.7 & 9.98 &  & X1.5 & 16.92 \\
        48 & 85 & 1989-04-10 21:15 & 1989-04-18 01:45 & 1.01E+09 & -1.91 & -6.29 & 8.29 & 36.23 & 2.42E+08 & 0.14 & 0.12 & 0.58 & 4.82 &  & X3.5 & 9.20 \\
        49 & 87 & 1989-05-24 05:00 & 1989-05-29 10:00 & 4.00E+07 & -1.52 & -7.37 & 6.89 & 40.3 & 1.08E+07 & 0.16 & 0.37 & 0.52 & 6.43 &  & M5.7 & 74.86 \\
        50 & 89 & 1989-07-25 09:05 & 1989-07-26 17:45 & 1.50E+06 & -1.17 & -2.56 & 51.3 & 71.25 & 3.60E+05 & 0.11 & 0.02 & 9.25 & 29.34 & Yes & X2.6 & 9.67 \\
        51 & 90 & 1989-08-12 15:45 & 1989-08-15 15:45 & 1.96E+08 & -0.25 & -4.52 & 9.64 & 41.11 & 4.12E+07 & 0.12 & 0.02 & 0.55 & 2.85 & Yes & X2.6 & 33.06 \\
        52 & 91 & 1989-09-12 13:30 & 1989-09-16 01:30 & 6.43E+07 & -1.85 & -5.89 & 11.73 & 47.44 & 1.28E+07 & 0.11 & 0.16 & 0.75 & 6.31 &  & M5.3 & 2.51 \\
        53 & 92 & 1989-09-29 11:55 & 1989-10-10 05:20 & 3.42E+08 & -0.99 & -2.83 & 24.53 & 44.95 & 6.32E+07 & 0.1 & 0.02 & 2.93 & 10.04 & Yes & X9.8 & 3.79 \\
        54 & 93 & 1989-10-19 13:10 & 1989-11-09 16:50 & 9.14E+09 & -1.35 & -2.89 & 13.2 & 20.35 & 3.90E+09 & 0.3 & 0.02 & 4.26 & 15.74 & Yes & X13 & 7.21 \\
        55 & 94 & 1989-11-15 07:25 & 1989-11-16 11:45 & 6.45E+06 & -1.92 & -3.78 & 328.25 & 608.35 & 3.82E+05 & 0.02 & 0.43 & 19.94 & 203.04 & Yes & X3.2 & 10.21 \\
        56 & 95 & 1989-11-27 06:25 & 1989-12-05 09:05 & 5.86E+08 & -0.3 & -5.07 & 4.58 & 21.87 & 2.34E+08 & 0.28 & 0.02 & 0.46 & 2.76 &  & X2.6 & 80.62 \\
        57 & 96 & 1990-03-19 06:30 & 1990-03-22 01:40 & 6.23E+08 & -1.26 & -7.1 & 8.23 & 48.12 & 9.05E+07 & 0.11 & 0.94 & 0.67 & 10.6 &  & X1 & 231.21 \\
        58 & 99 & 1990-04-16 06:05 & 1990-04-23 07:15 & 1.89E+08 & -2.26 & -5.15 & 14.94 & 43.14 & 4.02E+07 & 0.12 & 0.15 & 1.45 & 10.31 &  & X1.5 & 9.23 \\
        59 & 100 & 1990-04-28 05:30 & 1990-04-29 16:40 & 4.65E+07 & -1.33 & -5.38 & 10.63 & 43.04 & 9.74E+06 & 0.12 & 0.12 & 0.74 & 5.35 &  & C1.2 & 1.04 \\
        60 & 102 & 1990-05-17 21:30 & 1990-05-24 19:00 & 1.22E+08 & -1.98 & -2.78 & 90.48 & 71.72 & 1.68E+07 & 0.07 & 0.02 & 18.41 & 66.98 & Yes & X5.5 & 3.98 \\
        61 & 105 & 1990-07-26 04:20 & 1990-07-30 01:35 & 9.48E+08 & -3.15 & -3.89 & 55.55 & 41.08 & 2.02E+08 & 0.12 & 0.04 & 19.49 & 102.6 &  & M2.3 & 31.68 \\
        62 & 106 & 1990-07-31 15:25 & 1990-08-06 12:05 & 6.18E+08 & -2.01 & -5.35 & 11.67 & 38.9 & 1.36E+08 & 0.13 & 0.1 & 0.96 & 6.69 &  & M4.4 & 0.06 \\
        63 & 108 & 1991-01-27 14:45 & 1991-02-02 19:30 & 1.52E+09 & -2.3 & -5.81 & 7.29 & 25.61 & 4.80E+08 & 0.21 & 0.06 & 0.82 & 6.02 &  & X1.3 & 10.45 \\
        64 & 109 & 1991-02-08 09:55 & 1991-02-09 17:05 & 2.99E+07 & -1.98 & -6.88 & 7.93 & 38.86 & 6.98E+06 & 0.14 & 0.21 & 0.56 & 5.61 &  &  & 42.07 \\
        65 & 110 & 1991-02-25 10:40 & 1991-02-27 01:55 & 8.08E+06 & -1.53 & -5.39 & 8.25 & 31.82 & 2.28E+06 & 0.18 & 0.12 & 0.85 & 6 &  & X1.2 & 101.56 \\
        66 & 111 & 1991-03-12 18:50 & 1991-03-13 22:25 & 1.66E+06 & -0.91 & -5.36 & 8.77 & 39.1 & 3.77E+05 & 0.13 & 0.14 & 0.58 & 4.3 &  &  & 15.14 \\
        67 & 113 & 1991-03-23 06:40 & 1991-03-31 14:30 & 5.97E+08 & -0.29 & -4.29 & 7.58 & 30.29 & 1.72E+08 & 0.18 & 0.03 & 0.65 & 3.25 &  & X9.4 & 56.71 \\
        68 & 114 & 1991-04-02 06:45 & 1991-04-10 01:15 & 1.17E+09 & -2.63 & -5.42 & 14.39 & 40.28 & 2.61E+08 & 0.13 & 0.13 & 1.48 & 10.84 &  & M6 & 8.28 \\
        69 & 115 & 1991-04-22 12:00 & 1991-04-24 15:00 & 1.62E+07 & -2.03 & -6.58 & 12.33 & 56.08 & 3.84E+06 & 0.14 & 0.23 & 1.27 & 11.68 &  &  & 111.89 \\
        70 & 116 & 1991-05-10 15:05 & 1991-05-15 10:50 & 3.65E+08 & -2.23 & -3.91 & 32.3 & 54.29 & 5.79E+07 & 0.08 & 0.02 & 3.56 & 17.63 &  & M8.2 & 24.48 \\
        71 & 118 & 1991-06-09 19:10 & 1991-06-15 01:30 & 1.73E+09 & -1.97 & -3.71 & 45.87 & 80.01 & 2.36E+08 & 0.06 & 0.02 & 4.07 & 18.95 & Yes & X12 & 5.14 \\
        72 & 119 & 1991-06-29 21:25 & 1991-07-13 05:40 & 1.05E+10 & -1.97 & -5.24 & 7.09 & 23.19 & 3.86E+09 & 0.25 & 0.03 & 0.94 & 6.16 &  & X1.9 & 59.03 \\
        73 & 120 & 1991-08-25 21:10 & 1991-08-30 22:30 & 3.70E+09 & -2.76 & -6.25 & 11.52 & 40.25 & 8.20E+08 & 0.13 & 0.13 & 0.94 & 7.94 &  & X2.1 & 6.81 \\
        74 & 122 & 1991-09-30 09:40 & 1991-10-03 15:40 & 5.67E+08 & -3.21 & -4.44 & 33.13 & 40.94 & 1.19E+08 & 0.12 & 0.03 & 6.47 & 37.99 &  & M7.3 & 47.14 \\
        75 & 124 & 1991-10-30 07:30 & 1991-10-31 20:05 & 7.33E+06 & -1.54 & -2.73 & 48.42 & 57.28 & 1.04E+06 & 0.07 & 0.02 & 7.03 & 24.24 &  & X2.5 & 14.75 \\
        76 & 126 & 1991-12-29 05:40 & 1991-12-30 08:30 & 5.37E+07 & -2.45 & -5.03 & 5.41 & 13.95 & 4.83E+07 & 0.75 & 0.09 & 2.37 & 15.69 &  &  & 2.83 \\
        77 & 127 & 1992-02-06 22:45 & 1992-02-10 00:30 & 1.06E+09 & -1.61 & -5.43 & 3.46 & 13.23 & 1.04E+09 & 0.83 & 0.07 & 1.09 & 7.6 &  & M4 & 0.75 \\
        78 & 128 & 1992-02-27 11:40 & 1992-02-28 15:05 & 1.22E+08 & -2.85 & -4.15 & 8.57 & 11.2 & 1.72E+08 & 1.29 & 0.06 & 11.42 & 64.54 &  & C2.6 & 0.61 \\
        79 & 130 & 1992-05-09 06:15 & 1992-05-13 20:15 & 5.03E+09 & -1.84 & -5.15 & 5.98 & 19.81 & 2.31E+09 & 0.33 & 0.03 & 0.99 & 6.41 &  & M7.4 & 60.54 \\
        80 & 131 & 1992-06-25 20:30 & 1992-07-01 23:25 & 4.66E+07 & -0.74 & -3.19 & 8.36 & 20.43 & 2.02E+07 & 0.3 & 0.02 & 1.73 & 6.74 & Yes & X3.9 & 24.01 \\
        81 & 132 & 1992-08-05 20:35 & 1992-08-08 09:05 & 3.85E+07 & -1.49 & -10.07 & 5.15 & 44.22 & 1.16E+07 & 0.19 & 0.9 & 0.36 & 8.18 &  & M4.8 & 97.80 \\
        82 & 133 & 1992-10-30 18:45 & 1992-11-01 21:20 & 9.78E+07 & -0.32 & -5 & 10.31 & 48.28 & 1.64E+07 & 0.09 & 0.02 & 0.44 & 2.55 & Yes & X1.7 & 41.53 \\
        83 & 134 & 1993-03-04 13:20 & 1993-03-05 22:30 & 5.54E+06 & -1.71 & -3.68 & 22.58 & 44.38 & 1.09E+06 & 0.11 & 0.04 & 2.8 & 12.98 &  & C8.1 & 22.10 \\
        84 & 135 & 1993-03-06 23:05 & 1993-03-09 20:10 & 1.64E+08 & -3.06 & -4.05 & 52.96 & 52.46 & 3.02E+07 & 0.1 & 0.18 & 11.98 & 67.66 &  & C1.9 & 8.50 \\
        85 & 136 & 1993-03-12 18:50 & 1993-03-14 15:05 & 3.84E+07 & -2.09 & -3.28 & 33.13 & 39.39 & 8.12E+06 & 0.12 & 0.03 & 7.07 & 30.19 &  & M7 & 10.02 \\
        86 & 137 & 1994-02-20 02:20 & 1994-02-22 21:20 & 2.14E+08 & 0.73 & -6.7 & 2.55 & 18.95 & 1.04E+08 & 0.35 & 0.03 & 0.2 & 1.55 &  & M4 & 30.65 \\
        87 & 138 & 1994-10-19 22:35 & 1994-10-21 12:05 & 1.51E+07 & -0.18 & -3.73 & 2.73 & 9.71 & 4.44E+07 & 2.9 & 0.03 & 2.84 & 13.59 &  & M3.2 & 2.88 \\
        88 & 140 & 1997-11-06 06:00 & 1997-11-10 19:40 & 1.07E+08 & -1.47 & -3.04 & 58.41 & 91.92 & 1.30E+07 & 0.05 & 0.03 & 5.49 & 20.77 & Yes & X9.4 & 2.11 \\
        89 & 141 & 1998-04-20 12:55 & 1998-04-26 15:05 & 2.84E+07 & -0.19 & -5.42 & 12.56 & 65.69 & 4.11E+06 & 0.09 & 0.73 & 0.92 & 2.71 &  & M1.4 & 43.89 \\
        90 & 143 & 1998-05-02 13:55 & 1998-05-04 23:05 & 1.33E+07 & -1.47 & -3.34 & 57.79 & 108.07 & 1.48E+06 & 0.05 & 0.03 & 4.23 & 17.44 & Yes & X1.1 & 10.79 \\
        91 & 144 & 1998-05-06 08:25 & 1998-05-08 00:20 & 1.53E+07 & -1.54 & -3.15 & 28.04 & 45.04 & 3.00E+06 & 0.11 & 0.03 & 4.16 & 16.52 & Yes & X2 & 6.27 \\
        92 & 145 & 1998-05-09 06:50 & 1998-05-11 04:10 & 3.42E+06 & -1.38 & -4.32 & 15.56 & 45.74 & 7.00E+05 & 0.11 & 0.16 & 1.41 & 8.59 &  &  & 5.57 \\
        93 & 146 & 1998-06-17 16:00 & 1998-06-18 23:55 & 1.37E+06 & -1.03 & -3.07 & 5.02 & 10.25 & 2.85E+06 & 2 & 0.06 & 6.42 & 25.88 &  &  & 0.19 \\
        94 & 147 & 1998-08-22 17:25 & 1998-08-31 20:00 & 2.73E+09 & -2.44 & -3.85 & 26.22 & 36.91 & 5.92E+08 & 0.12 & 0.02 & 4.64 & 23.16 & Yes & X1 & 33.13 \\
        95 & 149 & 1998-09-30 14:25 & 1998-10-04 04:20 & 2.94E+08 & -1.25 & -4.24 & 9.77 & 29.22 & 8.62E+07 & 0.18 & 0.03 & 1.14 & 5.93 &  & M2.8 & 29.84 \\
        96 & 152 & 1998-11-14 06:30 & 1998-11-17 15:55 & 1.93E+07 & -1.1 & -3.73 & 21.18 & 55.66 & 3.15E+06 & 0.08 & 0.04 & 1.62 & 7.42 &  & C1.7 & 11.96 \\
        97 & 153 & 1999-01-21 00:30 & 1999-01-26 02:40 & 2.41E+08 & -2.86 & -4.21 & 34.39 & 46.35 & 4.94E+07 & 0.11 & 0.17 & 6.63 & 38.87 &  & M5.2 & 18.65 \\
        98 & 156 & 1999-05-27 12:15 & 1999-05-28 14:40 & 1.31E+06 & -1.42 & -4.36 & 15.14 & 44.47 & 2.76E+05 & 0.12 & 0.16 & 1.41 & 8.7 &  &  & 12.42 \\
        99 & 157 & 1999-06-01 22:10 & 1999-06-07 15:05 & 1.50E+08 & -1.94 & -4.32 & 17.94 & 42.8 & 3.04E+07 & 0.11 & 0.04 & 1.91 & 10.37 &  & M3.9 & 40.25 \\
        100 & 158 & 2000-02-18 08:45 & 2000-02-19 11:55 & 1.26E+06 & -1.32 & -5 & 18.42 & 67.72 & 1.29E+05 & 0.07 & 0.31 & 1.67 & 6.11 &  & M1.3 & 65.29 \\
        101 & 161 & 2000-06-10 04:00 & 2000-06-12 21:10 & 3.01E+07 & -2 & -3.93 & 31.73 & 61.15 & 4.95E+06 & 0.08 & 0.05 & 3.39 & 17.01 &  & M5.2 & 19.90 \\
        102 & 163 & 2000-07-13 00:30 & 2000-07-23 19:20 & 2.97E+08 & -0.15 & -3.63 & 12.05 & 41.98 & 6.15E+07 & 0.11 & 0.02 & 0.84 & 3.5 & Yes & X5.7 & 8.90 \\
        103 & 164 & 2000-07-28 02:55 & 2000-07-30 09:05 & 7.71E+05 & 0.07 & -10.51 & 4.36 & 46.1 & 2.13E+05 & 0.17 & 0.99 & 0.22 & 6.51 &  & M1.5 & 88.37 \\
        104 & 165 & 2000-08-13 00:45 & 2000-08-15 03:30 & 1.05E+09 & -3.09 & -7.42 & 3.53 & 15.29 & 1.11E+09 & 0.91 & 0.83 & 1.27 & 13.94 &  &  & 0.00 \\
        105 & 166 & 2000-09-12 14:50 & 2000-09-18 01:25 & 1.70E+09 & -2 & -4.36 & 7.2 & 17.04 & 5.67E+08 & 0.28 & 0.08 & 1.45 & 8.13 &  & M1 & 21.58 \\
        106 & 168 & 2000-10-25 15:15 & 2000-10-27 19:10 & 9.81E+07 & -2.38 & -5.73 & 12.38 & 41.44 & 2.19E+07 & 0.13 & 0.18 & 1.11 & 8.95 &  & M2 & 26.81 \\
        107 & 170 & 2000-11-08 23:45 & 2000-11-15 17:00 & 6.74E+08 & -0.87 & -5.1 & 24.26 & 102.77 & 7.49E+07 & 0.05 & 0.03 & 0.8 & 4.82 &  & M7.4 & 14.95 \\
        108 & 171 & 2000-11-24 07:00 & 2000-11-25 20:10 & 1.04E+08 & -2.3 & -3.43 & 41.05 & 46.58 & 1.91E+07 & 0.1 & 0.04 & 8.41 & 37.84 &  & X2 & 20.94 \\
        109 & 176 & 2001-04-02 11:20 & 2001-04-07 08:20 & 1.70E+08 & -1.21 & -4.05 & 15.13 & 42.96 & 3.50E+07 & 0.11 & 0.04 & 1.35 & 6.71 & Yes & X20 & 19.41 \\
        110 & 178 & 2001-05-07 15:00 & 2001-05-09 19:10 & 1.53E+08 & -2.31 & -5.45 & 12.14 & 38.1 & 3.36E+07 & 0.13 & 0.13 & 1.11 & 8.14 &  &  & 14.00 \\
        111 & 179 & 2001-05-20 08:45 & 2001-05-21 14:25 & 2.49E+05 & -0.8 & -2.99 & 22.84 & 49.9 & 2.82E+05 & 0.51 & 0.13 & 10.24 & 38.24 &  & M6.4 & 15.69 \\
        112 & 182 & 2001-08-16 00:55 & 2001-08-26 04:40 & 2.69E+07 & -1.13 & -5.14 & 38.3 & 153.78 & 2.48E+06 & 0.04 & 0.04 & 1.26 & 7.95 &  &  & 34.27 \\
        113 & 183 & 2001-09-24 12:00 & 2001-10-12 03:05 & 2.91E+09 & -1.45 & -5.52 & 21.52 & 87.64 & 3.55E+08 & 0.06 & 0.03 & 0.81 & 5.39 &  & X2.6 & 38.96 \\
        114 & 184 & 2001-10-19 04:55 & 2001-10-28 16:10 & 1.13E+08 & -2.52 & -3.01 & 148.55 & 73.58 & 1.82E+07 & 0.08 & 0.05 & 56.48 & 231.48 &  & X1 & 15.29 \\
        115 & 185 & 2001-11-05 16:55 & 2001-11-12 20:05 & 8.17E+08 & -0.73 & -4.8 & 15.86 & 64.49 & 1.20E+08 & 0.07 & 0.02 & 0.67 & 3.76 & Yes & X1 & 18.94 \\
        116 & 186 & 2001-11-17 19:55 & 2001-11-30 13:00 & 7.58E+08 & -0.46 & -5.08 & 7.77 & 35.84 & 1.81E+08 & 0.14 & 0.03 & 0.49 & 2.94 &  & M9.9 & 59.82 \\
        117 & 188 & 2001-12-26 05:55 & 2002-01-09 07:00 & 1.20E+09 & -2.23 & -3.45 & 58 & 70.41 & 1.60E+08 & 0.06 & 0.02 & 7.64 & 33.95 & Yes & M7.1 & 4.59 \\
        118 & 189 & 2002-01-10 09:55 & 2002-01-18 18:35 & 8.55E+08 & -2.4 & -4.96 & 15.84 & 40.58 & 1.77E+08 & 0.12 & 0.12 & 1.7 & 11.32 &  & C9 & 2.62 \\
        119 & 191 & 2002-04-17 11:30 & 2002-04-28 13:35 & 3.81E+08 & -1.14 & -7.03 & 23.15 & 136.35 & 3.90E+07 & 0.04 & 0.05 & 0.61 & 5.34 &  & X1.5 & 61.07 \\
        120 & 192 & 2002-05-22 07:50 & 2002-05-25 00:15 & 6.17E+09 & -3.45 & -4.55 & 44.37 & 48.82 & 1.14E+09 & 0.1 & 0.16 & 10.17 & 64.45 &  & C5 & 0.24 \\
        121 & 193 & 2002-07-07 14:00 & 2002-07-09 13:55 & 1.67E+06 & -0.83 & -4.26 & 11.97 & 41 & 3.64E+05 & 0.12 & 0.12 & 0.98 & 5.6 &  & M1 & 0.66 \\
        122 & 194 & 2002-07-16 13:20 & 2002-07-30 16:30 & 5.09E+09 & -3.05 & -4.46 & 32.46 & 45.49 & 1.02E+09 & 0.11 & 0.15 & 6.17 & 37.93 &  & X3 & 1.55 \\
        123 & 197 & 2002-08-22 03:10 & 2002-08-27 23:10 & 5.75E+08 & -2.22 & -3.21 & 70.74 & 70.36 & 7.62E+07 & 0.06 & 0.02 & 11.44 & 47.89 & Yes & X3.1 & 6.06 \\
        124 & 200 & 2002-11-09 17:10 & 2002-11-11 21:15 & 1.39E+08 & -1.31 & -4.81 & 7.65 & 26.73 & 4.25E+07 & 0.2 & 0.06 & 0.8 & 4.81 &  & M4.6 & 0.46 \\
        125 & 205 & 2003-10-28 03:30 & 2003-10-29 16:50 & 9.35E+08 & -0.73 & -4.18 & 15.31 & 52.8 & 1.61E+08 & 0.09 & 0.02 & 0.88 & 4.3 & Yes & X17 & 16.71 \\
        126 & 207 & 2003-12-02 12:50 & 2003-12-06 06:05 & 5.12E+07 & -0.63 & -9.87 & 4.41 & 40.73 & 1.33E+07 & 0.15 & 0.31 & 0.19 & 3.15 &  & C7 & 59.53 \\
        127 & 208 & 2004-04-11 06:45 & 2004-04-12 23:05 & 2.70E+09 & -3.51 & -5 & 24.91 & 37.08 & 6.08E+08 & 0.13 & 0.12 & 4.98 & 33.86 &  & C9.6 & 12.40 \\
        128 & 209 & 2004-07-23 15:05 & 2004-07-28 18:10 & 1.63E+09 & -2.29 & -5.37 & 11.27 & 34.78 & 4.02E+08 & 0.15 & 0.09 & 1.14 & 7.96 &  & M1.2 & 0.57 \\
        129 & 210 & 2004-07-31 20:55 & 2004-08-02 14:45 & 1.27E+06 & -0.33 & -9.67 & 4.66 & 43.55 & 3.77E+05 & 0.18 & 0.76 & 0.28 & 6.17 &  &  & 95.19 \\
        130 & 211 & 2004-09-13 19:55 & 2004-09-17 18:15 & 4.49E+09 & -2.38 & -6.29 & 7.2 & 28.15 & 1.38E+09 & 0.2 & 0.08 & 0.72 & 5.75 &  & M4.8 & 11.18 \\
        131 & 213 & 2004-11-01 06:10 & 2004-11-02 20:15 & 1.21E+06 & -0.63 & -3.69 & 14.62 & 44.81 & 2.35E+05 & 0.11 & 0.04 & 1.17 & 5.19 &  & M1.1 & 22.58 \\
        132 & 214 & 2004-11-07 02:50 & 2004-11-09 02:45 & 6.63E+09 & -3.09 & -6.7 & 26.75 & 96.34 & 7.76E+08 & 0.06 & 0.13 & 1.42 & 13.72 &  & X2 & 64.63 \\
        133 & 216 & 2005-01-20 01:00 & 2005-01-21 00:00 & 5.07E+07 & -1.01 & -2.57 & 47.53 & 74.14 & 6.61E+06 & 0.06 & 0.02 & 4.75 & 14.88 & Yes & X7.1 & 1.78 \\
        134 & 218 & 2005-05-13 21:00 & 2005-05-17 12:30 & 9.05E+08 & -0.69 & -5.78 & 3.65 & 18.53 & 4.69E+08 & 0.38 & 0.06 & 0.44 & 3.12 &  & M8 & 15.66 \\
        135 & 219 & 2005-06-16 20:50 & 2005-06-18 05:55 & 1.78E+06 & -0.83 & -3.21 & 23.33 & 55.58 & 2.85E+05 & 0.08 & 0.04 & 1.92 & 7.49 &  & M4 & 6.19 \\
        136 & 220 & 2005-07-13 18:10 & 2005-07-20 01:55 & 1.40E+08 & -1.42 & -4.74 & 11.79 & 39.15 & 3.06E+07 & 0.13 & 0.1 & 0.98 & 6.06 &  & M5 & 0.53 \\
        137 & 221 & 2005-07-26 23:20 & 2005-08-04 21:40 & 7.37E+08 & -2.19 & -4.64 & 14.99 & 36.77 & 1.71E+08 & 0.14 & 0.1 & 1.76 & 10.72 &  & M3.7 & 1.10 \\
        138 & 223 & 2005-09-07 23:25 & 2005-09-16 22:55 & 3.06E+10 & -2.97 & -5.37 & 155.41 & 371.88 & 2.02E+09 & 0.02 & 0.16 & 8.65 & 73.09 &  & X17 & 32.98 \\
        139 & 224 & 2006-12-05 17:35 & 2006-12-11 08:00 & 3.44E+09 & -2.07 & -4.71 & 38.62 & 102.26 & 3.79E+08 & 0.05 & 0.04 & 2.13 & 12.46 & Yes & X9 & 38.34 \\
        140 & 226 & 2010-08-18 08:50 & 2010-08-19 11:55 & 4.38E+06 & -1.58 & -3.63 & 5.88 & 12.06 & 5.04E+06 & 1.02 & 0.07 & 4.03 & 19.25 &  &  & 7.02 \\
        141 & 227 & 2011-03-07 23:15 & 2011-03-12 07:20 & 1.71E+08 & -1.71 & -5.04 & 10.12 & 33.69 & 4.36E+07 & 0.16 & 0.1 & 1.01 & 6.59 &  & M3.7 & 3.94 \\
        142 & 229 & 2011-06-05 20:05 & 2011-06-12 05:30 & 1.85E+07 & -1.41 & -3.2 & 32.25 & 57.75 & 3.03E+06 & 0.08 & 0.04 & 3.59 & 14.36 &  & M2.5 & 6.82 \\
        143 & 230 & 2011-08-04 04:35 & 2011-08-07 11:30 & 3.06E+08 & -2 & -3.91 & 17.11 & 32.72 & 7.81E+07 & 0.15 & 0.03 & 2.76 & 13.74 &  & M9.3 & 26.21 \\
        144 & 231 & 2011-08-09 08:00 & 2011-08-10 16:20 & 3.04E+06 & -1.37 & -3.21 & 29.15 & 53.64 & 5.26E+05 & 0.09 & 0.04 & 3.28 & 13.14 &  & X6.9 & 11.25 \\
        145 & 232 & 2011-09-06 23:45 & 2011-09-08 13:35 & 6.89E+05 & -1.03 & -3.74 & 22.54 & 61.15 & 1.11E+05 & 0.08 & 0.05 & 1.72 & 8.03 &  & C1.4 & 29.00 \\
        146 & 236 & 2011-11-26 09:00 & 2011-11-30 00:20 & 6.69E+08 & -1.96 & -5.36 & 7.47 & 25.41 & 2.21E+08 & 0.22 & 0.07 & 0.89 & 6.07 &  & C1 & 10.02 \\
        147 & 237 & 2012-01-26 11:25 & 2012-02-03 03:55 & 7.95E+07 & -0.66 & -4.1 & 11.91 & 40.98 & 1.60E+07 & 0.11 & 0.02 & 0.81 & 3.9 &  & X1.7 & 21.52 \\
        148 & 239 & 2012-03-05 00:30 & 2012-03-17 02:25 & 4.22E+09 & -1.88 & -4.22 & 42.74 & 99.9 & 4.79E+08 & 0.05 & 0.03 & 2.59 & 13.46 &  & X5.4 & 16.89 \\
        149 & 240 & 2012-05-17 02:05 & 2012-05-20 15:15 & 1.97E+08 & -2.17 & -2.91 & 110.62 & 81.23 & 2.58E+07 & 0.06 & 0.03 & 23 & 88.41 & Yes & M5.7 & 3.99 \\
        150 & 241 & 2012-05-26 23:55 & 2012-05-29 04:55 & 1.59E+07 & -0.91 & -4.11 & 4.35 & 13.89 & 1.40E+07 & 0.73 & 0.08 & 1.48 & 7.7 &  & C1.1 & 2.79 \\
        151 & 242 & 2012-06-16 09:45 & 2012-06-17 18:20 & 4.15E+08 & -2.99 & -4.34 & 9.22 & 12.39 & 4.49E+08 & 0.94 & 0.07 & 8.95 & 52.63 &  & M1.9 & 2.32 \\
        152 & 244 & 2012-07-12 18:05 & 2012-07-13 13:00 & 4.63E+06 & -0.04 & -5.54 & 4.67 & 25.66 & 1.53E+06 & 0.22 & 0.07 & 0.34 & 2.33 &  & X1.4 & 15.33 \\
        153 & 245 & 2012-07-17 15:50 & 2012-07-26 18:00 & 3.20E+08 & -1.96 & -5.24 & 25.24 & 82.86 & 4.06E+07 & 0.06 & 0.09 & 1.39 & 9.84 &  & M1.8 & 47.88 \\
        154 & 247 & 2012-09-28 01:20 & 2012-10-01 03:35 & 9.89E+06 & -1.5 & -4.33 & 14.6 & 41.29 & 2.16E+06 & 0.12 & 0.13 & 1.47 & 8.54 &  & C3.7 & 2.97 \\
        155 & 249 & 2013-03-15 19:45 & 2013-03-18 12:55 & 2.17E+08 & -1.84 & -9.62 & 5.41 & 42.15 & 6.27E+07 & 0.18 & 0.62 & 0.38 & 7.01 &  & M1.1 & 89.54 \\
        156 & 250 & 2013-04-11 09:25 & 2013-04-14 17:00 & 2.84E+07 & -1.35 & -3.33 & 18 & 35.55 & 6.59E+06 & 0.14 & 0.03 & 2.52 & 10.42 &  & M6.5 & 12.72 \\
        157 & 251 & 2013-05-14 07:20 & 2013-05-21 15:20 & 3.67E+08 & -1.77 & -5.61 & 8.48 & 32.58 & 9.91E+07 & 0.17 & 0.11 & 0.8 & 5.82 &  & X1.2 & 13.12 \\
        158 & 252 & 2013-05-22 13:45 & 2013-05-25 23:35 & 9.07E+07 & -0.02 & -4.21 & 5.41 & 22.63 & 3.53E+07 & 0.26 & 0.03 & 0.59 & 2.95 &  & M5 & 40.23 \\
        159 & 255 & 2014-02-25 10:15 & 2014-03-05 19:00 & 1.74E+08 & -0.86 & -3.54 & 5.46 & 14.64 & 1.30E+08 & 0.6 & 0.02 & 1.8 & 8.02 &  & X4.9 & 20.66 \\
        160 & 256 & 2014-04-18 14:30 & 2014-04-21 08:55 & 3.34E+09 & -3.11 & -4.15 & 22.61 & 23.52 & 1.19E+09 & 0.24 & 0.03 & 9.2 & 51.28 &  & M7.3 & 32.65 \\
        161 & 260 & 2015-06-18 05:05 & 2015-07-02 17:05 & 5.02E+09 & -2 & -6.43 & 6.32 & 27.98 & 1.55E+09 & 0.2 & 0.09 & 0.57 & 4.65 &  & M2 & 22.17 \\
        162 & 261 & 2016-01-02 01:05 & 2016-01-03 12:55 & 2.68E+08 & -3.04 & -5 & 19.83 & 38.99 & 5.88E+07 & 0.13 & 0.15 & 3.04 & 20.84 &  & M2.4 & 20.94 \\
        163 & 262 & 2017-07-14 04:50 & 2017-07-16 13:55 & 3.36E+08 & -2.34 & -6.52 & 9.54 & 39.84 & 8.13E+07 & 0.14 & 0.24 & 0.81 & 7.8 &  & M2.4 & 46.82 \\
        164 & 263 & 2017-09-10 05:30 & 2017-09-15 12:40 & 4.06E+08 & -1.28 & -3.56 & 29.56 & 67.13 & 5.92E+08 & 0.07 & 0.02 & 2.25 & 9.68 & Yes & X8 & 6.83\\
\enddata
\end{deluxetable*}}
\vspace{-26pt}
\tablenotetext{1}{Number in SEPEM reference event list.}
\tablenotetext{2}{The normalization coefficient for Equation \ref{eqn:Band_function} and its uncertainty are expressed in units of $[\#/cm^2/sr/MeV]$.}
\tablenotetext{3}{The $E_0$ for Equation \ref{eqn:Band_function} and its uncertainty are expressed in units of $[MeV]$.}
\tablenotetext{4}{The break-point energy for Equation \ref{eqn:Band_function} and its uncertainty are expressed in units of $[MeV]$..}
\end{longrotatetable}



\clearpage
\listofchanges

\bibliography{main}{}

\end{document}